\documentclass[showpacs,aps,twocolumn,prd,longbibliography]{revtex4-2}
\usepackage{epsfig}
\usepackage{graphicx}
\usepackage{amsmath,amssymb,amsfonts}
\usepackage{array}
\usepackage{url}
\usepackage{multirow}
\usepackage{float}
\usepackage{lineno}
\usepackage{xspace}
\usepackage{ulem}
\usepackage{bm}
\usepackage{rotating}

\newcommand{\bef}{\begin{figure}}
\newcommand{\eef}{\end{figure}}
\newcommand{\bc}{\begin{center}}
\newcommand{\ec}{\end{center}}

\newcommand{\be}{\begin{equation}}
\newcommand{\ee}{\end{equation}}
\newcommand{\bea}{\begin{eqnarray}}
\newcommand{\eea}{\end{eqnarray}}
\newcommand{\pT}{$p_{\rm{T}}$}
\newcommand{\DCAXY}{$\rm{DCA_{XY}}$}
\newcommand{\DCAZ}{$\rm{DCA_{Z}}$}

\def\ba{\begin{eqnarray}}
\def\ea{\end{eqnarray}}

\usepackage[usenames,dvipsnames]{color}
\definecolor{darkblue}{RGB}{0,0,196}
\usepackage[colorlinks=true,linkcolor=darkblue,citecolor=darkblue,urlcolor=darkblue]{hyperref}

\begin{document}
\title{\textbf \LARGE {Machine learning driven identification of heavy flavor decay leptons in proton-proton collisions at the Large Hadron Collider}}

\author{Raghunath Sahoo}
\email[Corresponding Author: ]{Raghunath.Sahoo@cern.ch}
\author{Kangkan Goswami}
\author{Suraj Prasad}

\affiliation{Department of Physics, Indian Institute of Technology Indore, Simrol, Indore 453552, India}

\begin{abstract}
    The study of heavy-flavor hadrons is topical in the era of precision measurements, which is useful to test theories based on pQCD. The heavy-flavor hadrons are produced initially during heavy-ion or hadronic collisions and are one of the best probes to understand the initial stages of the collisions as well as the system evolution. In experiments, the heavy-flavor sectors are studied directly via their decay to different hadrons or di-leptons or via their semi-leptonic decay, which is accompanied by additional neutrinos. However, their measurement in experiments is resource-intensive and requires input from different Monte-Carlo event generators. In this study, we provide an independent method based on Machine Learning algorithms to separate such leptons coming from heavy-flavor semi-leptonic decays. We use PYTHIA8 to generate events for this study, which gives a good qualitative and quantitative description of heavy-flavor production in $pp$ collisions. We use the XGBoost model for this study, which is trained with $pp$ collisions at $\sqrt{s}=13.6$~TeV. We use \DCAXY, \DCAZ~and pseudo-rapidity as the input to the machine. The ML model provides an accuracy of 98\% for heavy-flavor decay electrons and almost 100\% for heavy-flavor decay muons. 
\end{abstract}
\date{\today}
\maketitle
\section{Introduction}
In an attempt to explore and understand the elementary building blocks of the universe, ultra-relativistic heavy ions are collided at facilities like the Large Hadron Collider (LHC) at CERN and the Relativistic Heavy-Ion Collider (RHIC) at BNL. These collisions allow us to access a deconfined and thermalized matter of quarks and gluons, namely, Quark Gluon Plasma (QGP). However, this transient state of matter can't be detected directly as the produced partons hadronize within the timescale of $10^{-23}$ seconds. We often make use of probes, such as heavy-flavour hadrons consisting of heavy quarks (HQs), to understand the interactions within the deconfined medium. The heavy quarks are produced in the initial hard scattering, and their study can act as a good testing ground for perturbative Quantum Chromodynamics (pQCD). These HQs traverse the whole deconfined medium to form open-heavy flavor hadrons (or Quarkonium states) at the deconfinement phase boundary. This makes them an excellent probe for studying the deconfined medium and understanding the dynamics of the hadronic phase.

In experiments, to understand the production mechanism of heavy-flavor hadrons, the self-normalized yield of the heavy-flavor hadron is studied as a function of self-normalized charged particle multiplicity. This offers valuable insights into the underlying partonic dynamics and the interplay between hard and soft processes involved in particle production. In experiments, for $pp$ and $p-Pb$ collision, self-normalized yields of $J/\psi$~\cite{ALICE:2020msa, STAR:2018smh, ALICE:2017wet, ALICE:2020eji}, $D$ meson~\cite{ALICE:2015ikl}, $\Upsilon$(nS)~\cite{CMS:2013jsu} have been estimated at different center of mass energies. Moreover, the angular correlation between heavy flavor hadrons and charged particles has been studied in experiments, as it is sensitive to the production mechanism and hadronization of heavy quarks~\cite{Norrbin:2000zc}. In principle, the near-side peak is due to fragmentation of the same parton that produced the trigger particle. However, the away-side peak is induced due to the fragmentation of the other partons. The azimuthal correlation of prompt D mesons with charged particles has been studied at ALICE~\cite{ALICE:2021kpy, ALICE:2019oyn} for $pp$ collisions at $\sqrt{s}=13$ TeV and for $pp$ and $p-Pb$ collisions at $\sqrt{s_{\rm NN}}=5.02$ TeV, respectively. 

Although these studies are often done via hadronic decays of the D meson. The open-heavy flavors (B \& D) hadrons often decay via a semi-leptonic channel, $B, D \rightarrow l^{\pm} + \nu_{l} + X$, with a branching ratio of 0.11~\cite{ParticleDataGroup:2024cfk}. These semi-leptonic decays can't be reconstructed due to the neutrino carrying some of the parent's momentum. Hence, the preferred way of measuring these heavy-flavor decay leptons is to subtract the non-heavy-flavor leptons from the inclusive lepton spectra~\cite{ALICE:2020hdw, STAR:2014yia, ALICE:2016ywm, ALICE:2023xiu, ALICE:2019rmo, ALICE:2019nuy}. Owing to the large cross-section, this decay channel is widely explored in experiments to study heavy-flavor hadrons. The heavy flavor decay electrons are used in studying the momentum anisotropy in experiments like STAR and ALICE across various center-of-mass energies, from $\sqrt{s_{\rm NN}}=39$ GeV Au-Au collisions in STAR to $\sqrt{s_{\rm NN}}=5.02$ TeV Pb-Pb collisions in ALICE~\cite{STAR:2014yia, ALICE:2016ywm, ALICE:2020hdw}. Moreover, recently, production and nuclear modification factor of heavy flavor decay leptons have been estimated in ALICE~\cite{ALICE:2023xiu, ALICE:2019rmo, ALICE:2019nuy, ALICE:2016mpw}. However, most of the experimental measurements of heavy-flavor decay electrons heavily rely on the input from the Monte Carlo models. Moreover, a track-level identification of heavy-flavor decay leptons using the traditional methods is not possible.

On the other hand, machine learning techniques have been employed in high-energy physics for several decades~\cite{Bowser-Chao:1992giy, Chiappetta:1993zv}. They have been successfully applied to a wide range of studies, including jet measurements and tagging ~\cite{ALICE:2023waz, ATLAS:2020iwa, Hassan:2025yhp, Draguet:2025efc}, particle reconstruction~\cite{CMS:2020poo, Graczykowski:2022zae, Tamir:2023aiz}, event shape estimation~\cite{Basak:2025idd, Ortiz:2020rwg, Prasad:2025yfj, Mallick:2021wop}, impact parameter estimation~\cite{Bass:1996ez, Mallick:2021wop, Zhang:2021zxd}, and the measurement of flow coefficients~\cite{Mallick:2022alr, Mallick:2023vgi, Hirvonen:2023lqy}. Furthermore, ML algorithms have been successfully applied in designing beam-pipe material in particle accelerators~\cite{Singh:2024ihp}. More recently, classification algorithms have been effectively used to distinguish prompt and nonprompt charmed hadrons, like $J/\psi$ at forward and mid rapidity~\cite{Prasad:2023zdd}, $D^{0}$ meson~\cite{Goswami:2024xrx}, and $\lambda_{\rm{c}}$ baryon~\cite{Mallick:2025bsn}. These studies show the capability of machine learning models to study the heavy-flavour sector of particle production, with track-level identification of particle species and an independent way of separating them from the background particles. 

In this Letter, we attempt to separate the heavy flavor decay leptons from the inclusive lepton sample. We use a classification algorithm, specifically, the gradient boosting technique XGBoost, known for its high accuracy and efficiency.
We primarily take advantage of the displaced distance of closest approach (DCA) for weakly decaying heavy-flavor decay leptons. Along with DCA, in the XY plane (\DCAXY) and along the z-direction (\DCAZ),  we use the pseudo-rapidity of the leptons as well. Topological features such as DCA and decay length can be of great importance in separating daughter particles decaying via weak interaction. As a proof of concept, this study is performed using PYTHIA8-generated events, which offer a better handling and control of the $pp$ collisions with track-level truth information. It is ideal for evaluating the efficiency of the machine learning algorithm under ideal conditions. Furthermore, we estimate the self-normalized yield and azimuthal correlation and compare the results obtained from PYTHIA8 and XGBoost prediction. This comparison allows us to evaluate the model’s ability to reproduce physics observables when trained with track-level truth information. We train the model for $pp$ collisions at $\sqrt{s}=13$ TeV and test the XGB model for different center of mass energies, such as 900 GeV and 7 TeV. In Section~\ref{sec: Methodology}, we discuss the methodology of this work and in details about the machine learning algorithm. Section~\ref{sec: Results} presents the results and discussion, and finally, in Section~\ref{sec: Summary}, we conclude our work.

\section{Methodology}
\label{sec: Methodology}
This section presents a brief introduction to PYTHIA8 and XGBoost algorithms. Moreover, we discuss the background sources for the electrons and the training datasets.
\subsection{PYTHIA8}

PYTHIA8 is a Monte Carlo event generator used to simulate leptonic, hadronic, and heavy-ion collisions. PYTHIA8 was developed as a significant upgrade over PYTHIA6, with improved tuning to the LHC data \cite{Sjostrand:2007gs}. We have generated around $10^{9}$ events of proton-proton collisions at the center-of-mass energy of 13 TeV. 
To mimic an experiment-like environment, the {\tt Beams:allowVertexSpread = on} is set so that the vertex of two colliding beams would follow a normal distribution and not the origin instead. And to limit it, we used {\tt Beams:maxDevVertex=5} such that the maximum total deviation of the vertex from the origin is restricted. The {\tt ColourReconnection:reconnect=on} is set to allow a system to be merged with the other one, and the mode is set as {\tt ColourReconnection:mode=2} such that the calculations follow the Gluon Move Model. The tune 4C is used as {\tt Tune:pp=5} as it best aligns with the experimental results so far. Also, {\tt HardQCD:all=on} is a switch for all hard QCD 2 $\rightarrow$ 2 processes. \cite{Bierlich:2022pfr}
\begin{figure}[h!]
    \centering
    \includegraphics[width = 0.5\textwidth]{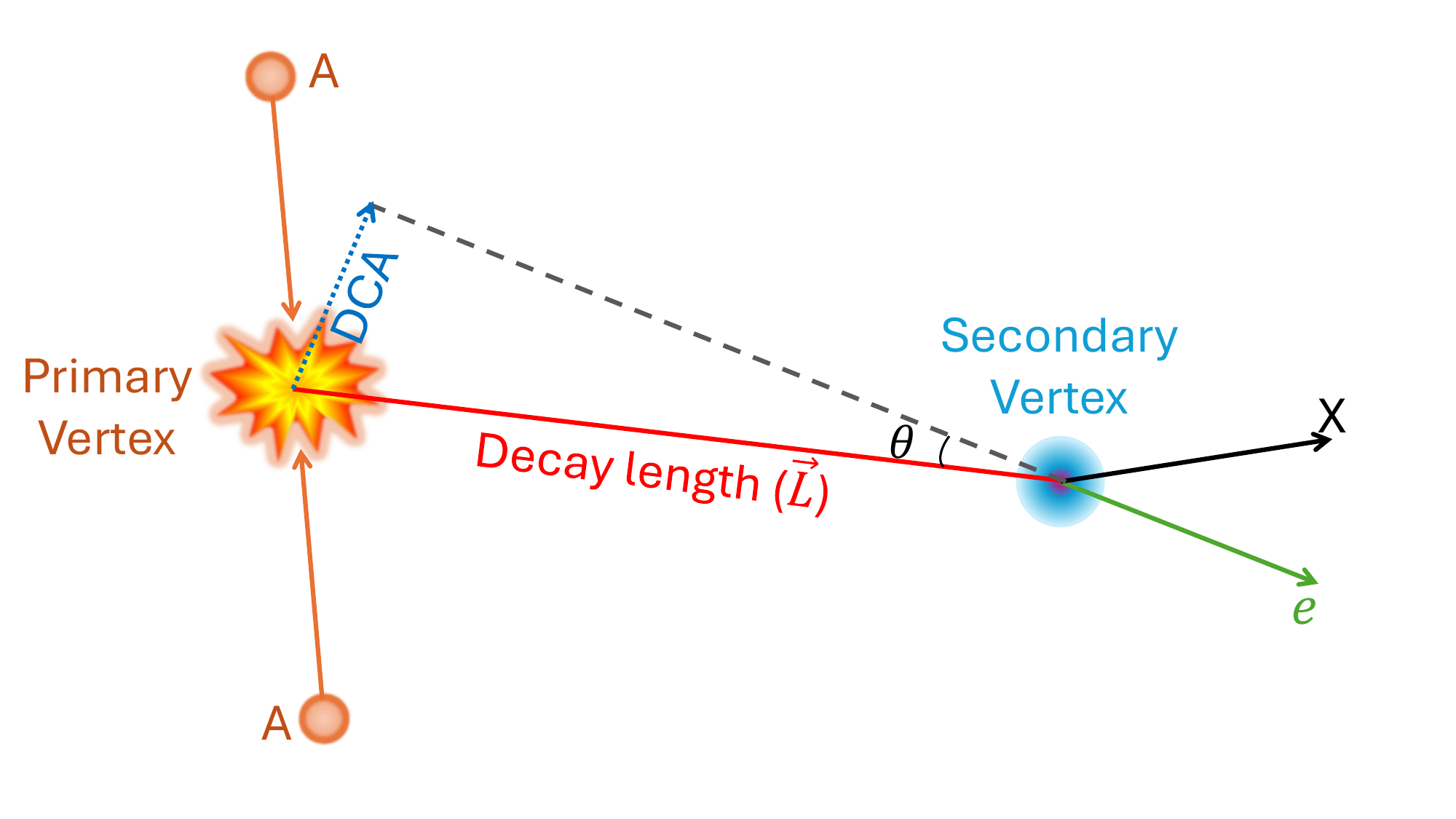} 
    \caption{Diagram representing the production topology of heavy flavor decay electrons.}
    \label{DCA_diagram}
\end{figure}


\begin{figure*}
    \centering
    \includegraphics[width = 0.42\textwidth]{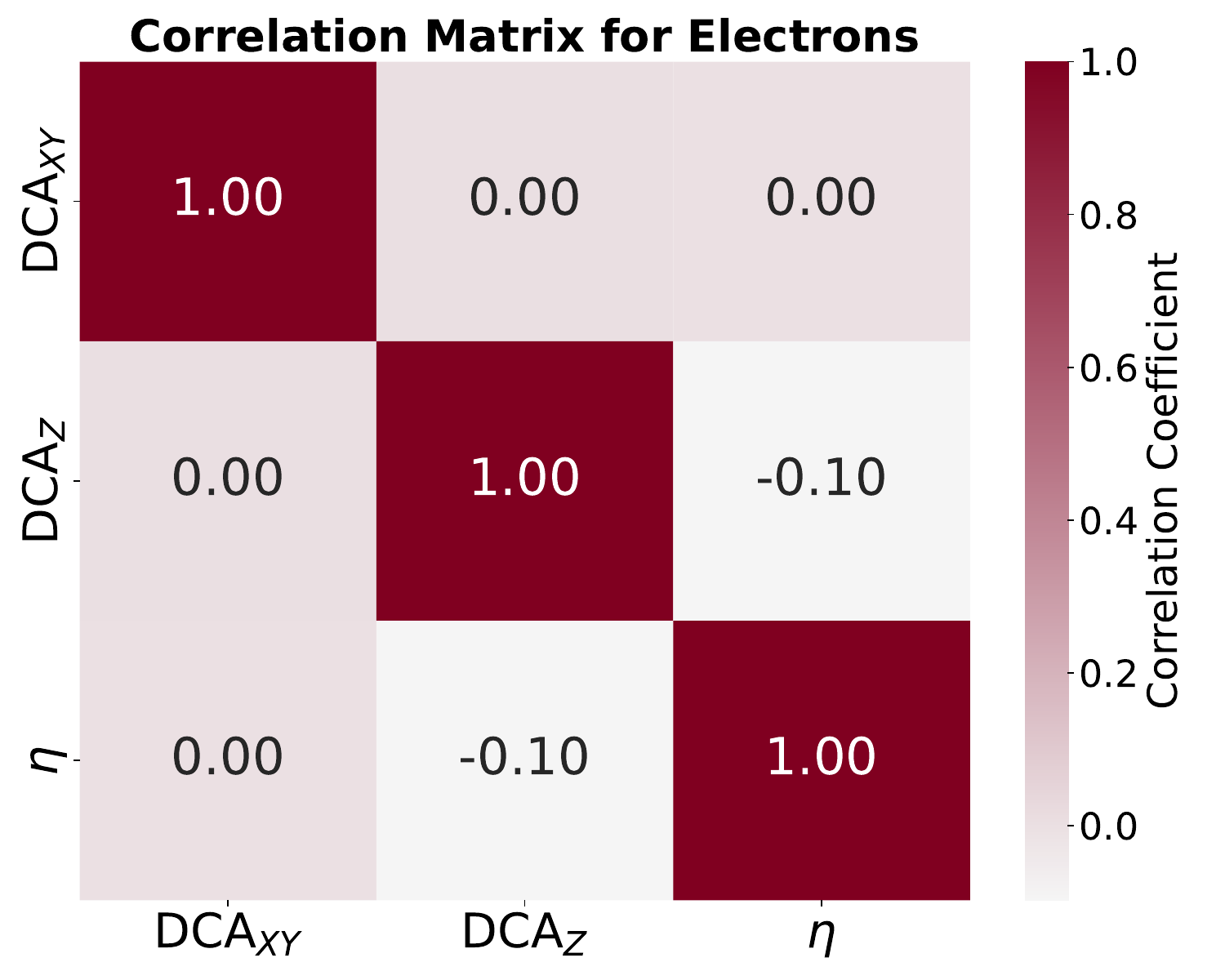} 
     \includegraphics[width = 0.42\textwidth]{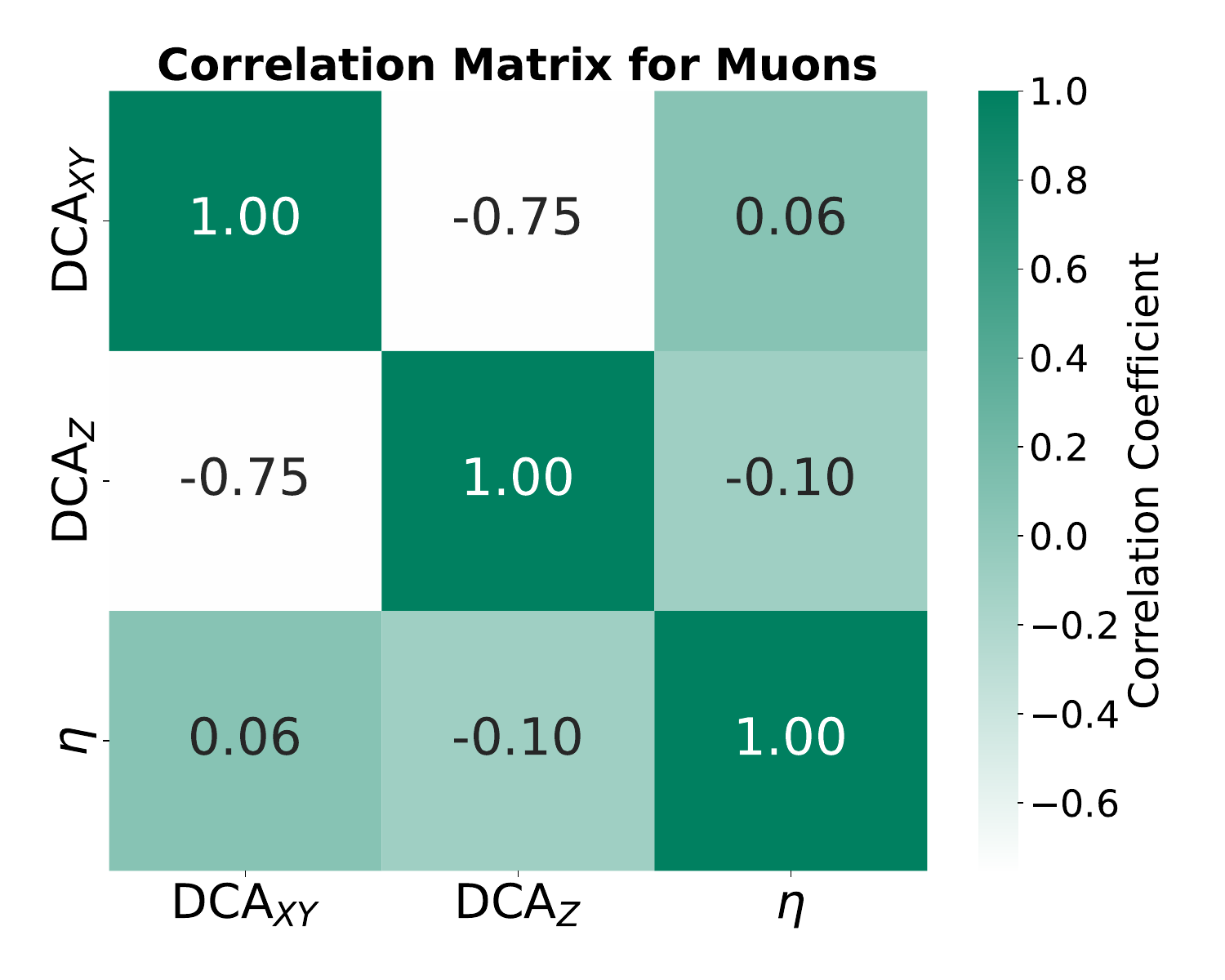} 
    \caption{Correlation matrix showing the Pearson's correlation coefficient among input variables for electrons (left) and muons (right).}
    \label{fig_Correlation_matrix}
\end{figure*}
\subsection{Machine Learning Model}

Decision trees are versatile supervised learning models used for both classification and regression problems. It works by recursively partitioning the data based on specific conditions that aim to best separate the classes. These trees can be understood structurally, as they resemble inverted trees with the root node representing the entire unclassified data. Meanwhile, the branches denote progressively classified subsets. A node that can not be further split is taken as a pure node and termed as \textit{leafnode}. The criteria often used to determine the splitting conditions include metrics such as \textit{entropy} and \textit{information gain}. A widely used method to combine such weak learners (shallow decision trees) is termed the Gradient Boosting Algorithm. One such algorithm is the eXtreme Gradient Boosting (XGBoost or XGB) algorithm, which incorporates regularization and parallel tree construction to achieve high predictive performance and computational scalability. The details of model parameters can be found in Refs.~\cite{xgb:2016, xgb:web}. XGB is shown to be quite robust to identify different topologically produced heavy-flavour hadrons such as J/$\psi$ and $D^{0}$ mesons from the combinatorial background as discussed in Refs.~\cite{Prasad:2023zdd, Goswami:2024xrx} with minimal hyperparameter tuning.

\subsection{Training Dataset}
\label{sec:training}
In this study, we attempt to separate the heavy flavor decay leptons from the total combinatorial background. The background comprises the particle contributions from the following processes~\cite{ALICE:2023xiu}:

\begin{enumerate}
    \item Photon conversions of the leptons, including Drell-Yan processes: ($\gamma\gamma\rightarrow l^+l^-$)
    \item Decays of light flavor hadrons: ($K_{e3}, \eta, \rho, \omega, \Lambda, \Sigma, \phi$)
    \item Dalitz decays: ($\eta\rightarrow\gamma l^+ l^-$)
    \item Decay of quarkonia into leptons: ($J/\psi\rightarrow l^+l^-$) 
\end{enumerate}

\begin{figure}
    \centering
    \includegraphics[width = 0.5\textwidth]{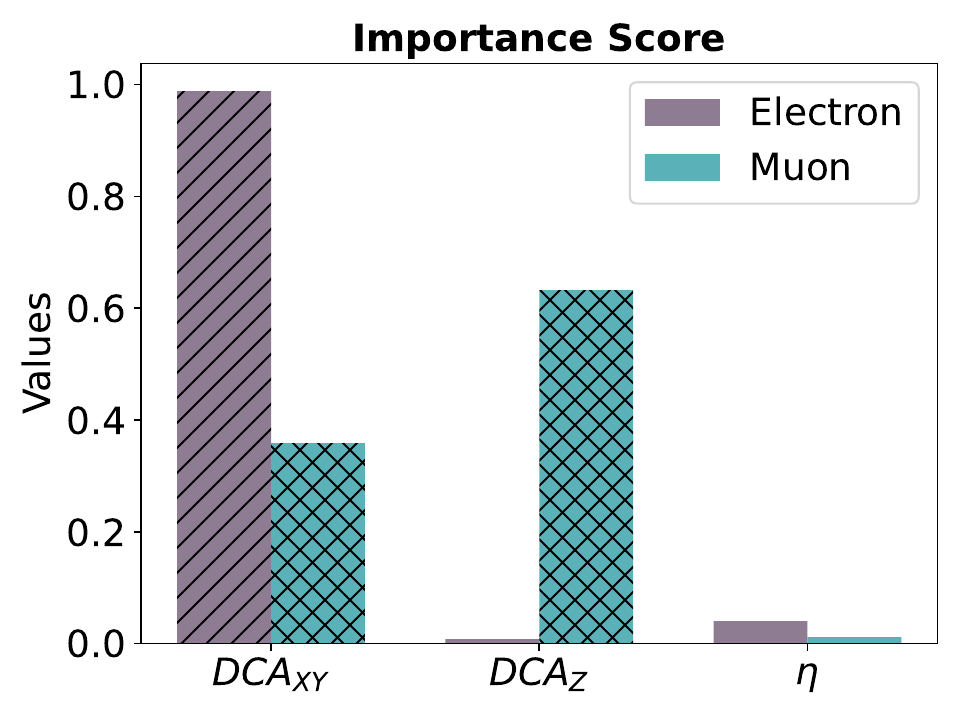} 
    \caption{Comparison of importance score of input parameters for electrons and muons in training the XGBoost algorithm.}
    \label{imp_score}
\end{figure}

We select events having $z$-component of the production vertex, $|V_z| \leq 100.0$, to avoid extreme deviation from the origin in an actual experimental setup. We keep the mid-rapidity window $|y|<0.8$ for electrons and forward rapidity $2.5<y<4.0$ for muons to match the experimental cuts as in ALICE. PYTHIA8 allows us to check the mother particles of the final state leptons, thus enabling us to tag the heavy flavor decay leptons and the background separately. We make use of this tag in the later stages of training the machine-learning algorithms.
\begin{figure}
    \centering
    \includegraphics[width = 0.5\textwidth]{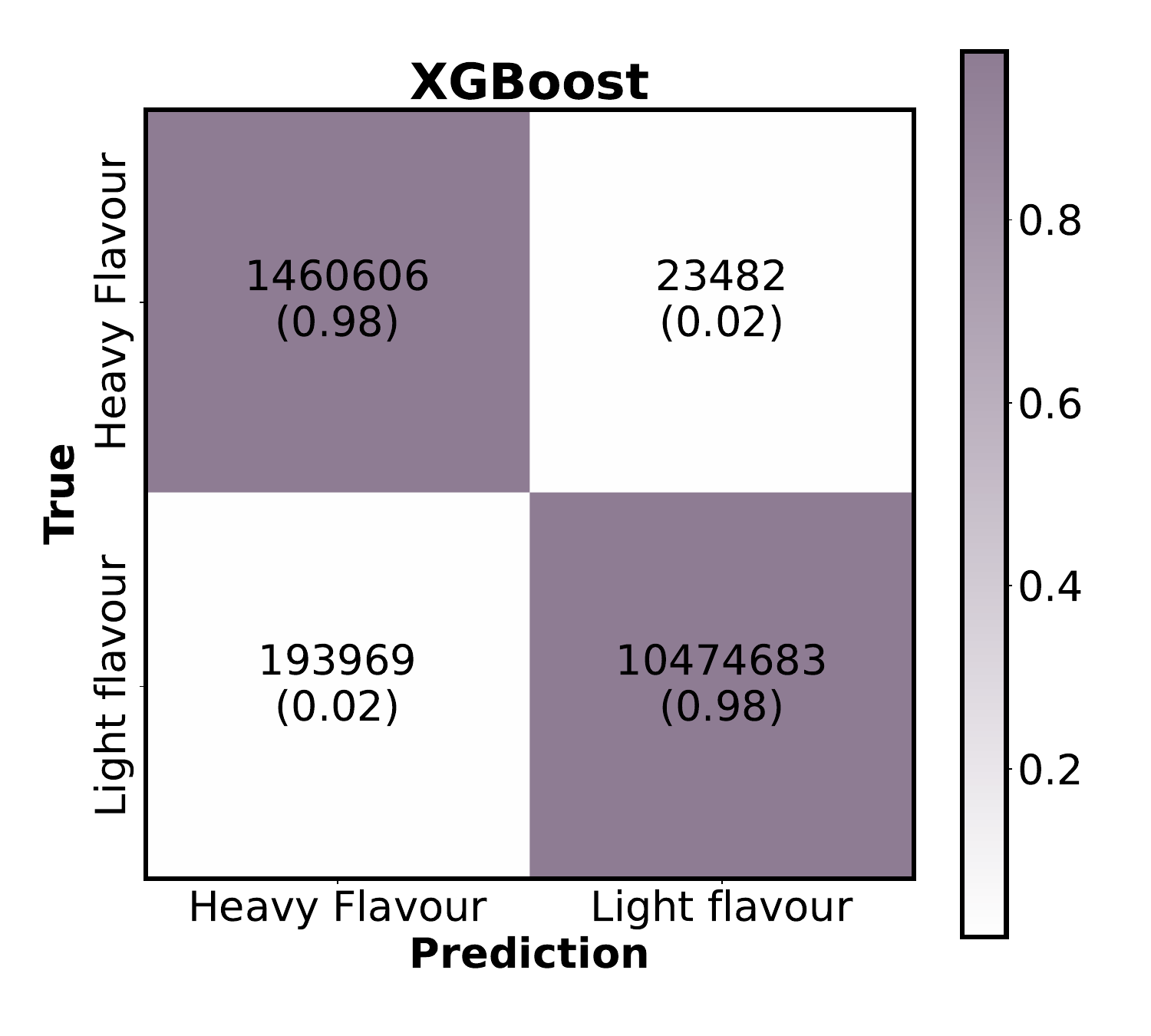} 
    \caption{Confusion Matrix for predicted value from XGBoost model for electrons coming from the decay of heavy-flavor and light hadron decays.}
    \label{fig_CM_electron}
\end{figure}

\begin{figure}
    \centering
    \includegraphics[width = 0.5\textwidth]{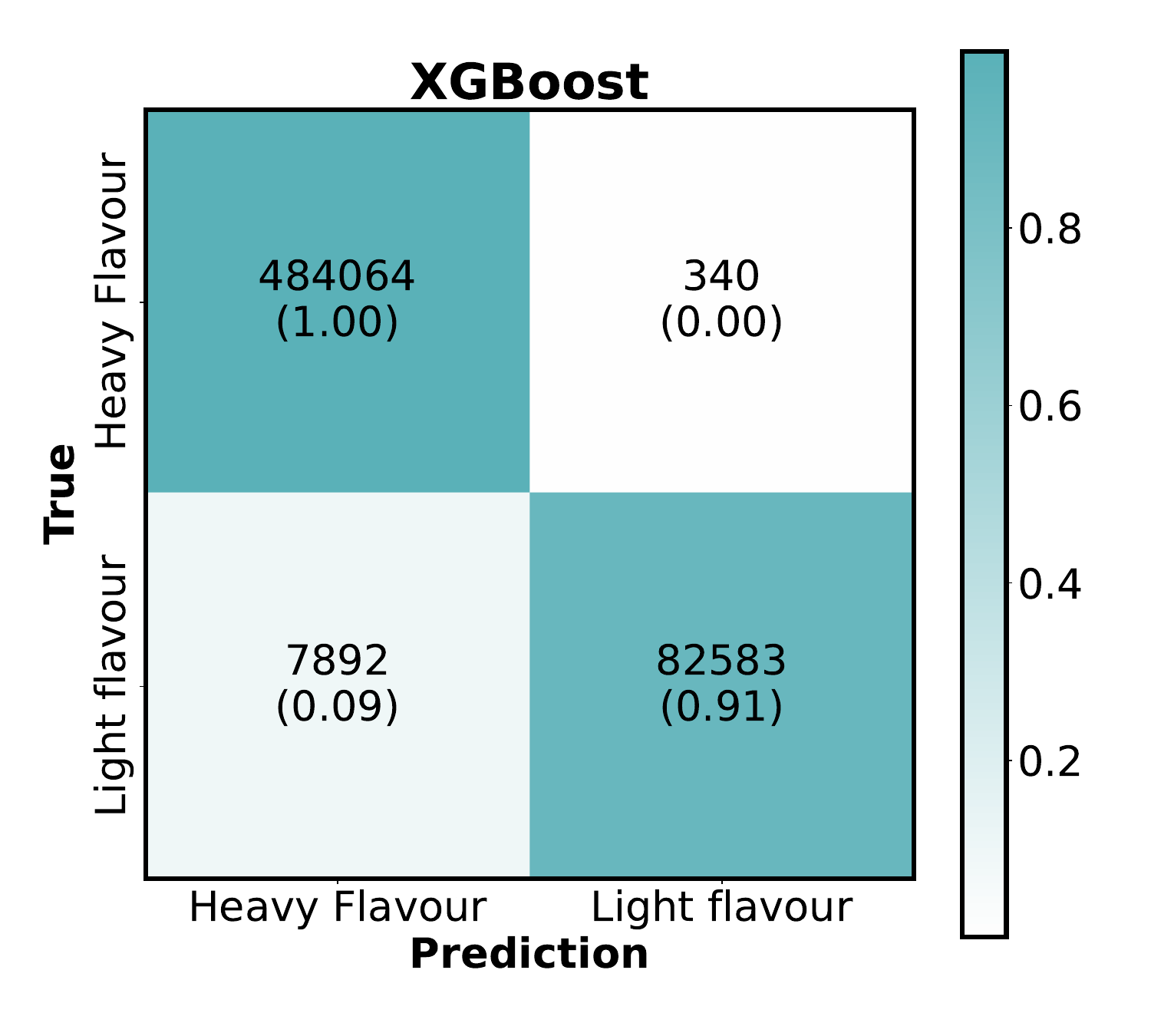} 
    \caption{Confusion Matrix for predicted value from XGBoost model for muons coming from the decay of heavy-flavor and light hadron decays.}
    \label{fig_CM_muon}
\end{figure}
Primarily, we aim to separate the electrons (or muons), decaying from heavy-flavor hadrons, from the background. For this purpose, we use the distance of closest approach~(\DCAXY~and \DCAZ) and pseudo-rapidity in different $p_{\rm{T}}$ of the leptons as the parameter space. The distance of closest approach is defined here as~\cite{STAR:2018zdy},
\begin{equation}
    {\rm DCA_{XY}} = |\vec{L}_{xy}|\cdot \sin{\theta},
\end{equation}
\begin{equation}
    {\rm DCA_{Z}} = |\vec{L}_{z}|\cdot \sin{\theta},
\end{equation}
where $\vec{L}$ is the vector representing the distance from the primary vertex to the decay vertex of the electron, while $\theta$, is the angle between $\vec{L}$ and the momentum vector $\vec{p}$ of the electron. Here, $|\vec{L}_{xy}|$ represents the projection of $\vec{L}$ in the XY plane and $|\vec{L}_{z}|$ is the component in the Z-direction. A pictorial representation of decay topology of the heavy flavor decay electrons is shown in Fig.~\ref{DCA_diagram}. The distribution of ${\rm DCA}_{\rm XY}$ and ${\rm DCA}_{\rm Z}$ for electrons and muons from PYTHIA8 is shown in Figs.~\ref{fig:DCA_electron} and \ref{fig:DCA_muon}, respectively in Appendix~\ref{App:B}. A detailed calculation of $\vec{L}$ can be found in Ref.~\cite{Prasad:2023zdd}.

In our study, we encounter a very high-class imbalance along with a limited feature space. The class imbalance stems from the fact that the production of heavy-flavor decay electrons (HFE) and heavy-flavor decay muons (HFM) is much less than that of their light-flavor counterparts. Thus, the leptons coming from the heavy-flavor decay are comparatively very small. This results in a huge class imbalance between the leptons coming from heavy and light flavors. To overcome this, we use over-sampling techniques such as Synthetic Minority Over-sampling Technique~(SMOTE), choose appropriate hyperparameters, and train the model for the different $p_{\rm{T}}$ bins. The performance of oversampling uisng SMOTE is shown in Figs.~\ref{fig:scatter-SMOTE} and \ref{fig:ratio-SMOTE}, in the appendix. Training the model $p_{\rm{T}}$ bin-wise allows us to carefully optimize the XGB model for different \pT~regions. This is due to the difference in the contribution from the light and heavy flavor sectors throughout the whole \pT~range. Additionally, we observe that the contribution from the light-flavor sector is maximum in the low $p_{\rm{T}}$ region. 

Moreover, to determine the architecture of the ML model, it is crucial to choose appropriate hyperparameters to achieve optimum results. To evaluate the performance of the XGB model, we make use of metrics such as precision, recall, and F1 score.
Precision can be understood as the fraction of selected leptons that truly come from heavy-flavor decays, i.e.,
$
\text{Precision} = TP/(TP + FP),
$
where $TP$ (true positives) is the number of correctly identified heavy-flavor decay leptons, and $FP$ (false positives) is the number of leptons incorrectly identified as heavy-flavor. Furthermore, recall is the fraction of all heavy-flavor decay leptons present in the data that are successfully identified, i.e.,  
$
\text{Recall} = TP/(TP + FN),
$  
where $FN$ (false negatives) is the number of heavy-flavor decay leptons that were missed by the identification. Additionally, the F1 score is a single metric that balances both precision and recall, defined as their harmonic mean, i.e.,  
$
F1 = 2 \times \frac{\text{Precision} \times \text{Recall}}{\text{Precision} + \text{Recall}}.
$
Since our goal is to maximize precision (to reduce the misidentification) and recall (to lower the misidentification of true heavy-flavor decay leptons), we take the F1 score as our target. However, there are various ways to tune the hyperparameters; we used Bayesian Optimization. It uses a probabilistic model to efficiently search the hyperparameter space based on prior evaluations. Compared to exhaustive methods like Grid Search, Bayesian Optimization is significantly more efficient in terms of computation and time.

The tuned hyperparameters are as follows,
\begin{itemize}
    \item \textit{n\_estimators}: It assigns the number of decision trees (or weak learners). Increasing this generally improves performance, but beyond a point, it can cause overfitting.
    \item \textit{learning\_rate}: It controls how quickly the model adapts during training. A high learning rate may lead to underfitting, while a very low one increases training time and the risk of overfitting.
    \item \textit{max\_depth}: Determines the maximum depth of each decision tree, i.e., the distance between the root and the leaf node, which affects its complexity. Deeper trees can model more intricate patterns, but are also more prone to overfitting and computationally expensive.
    \item \textit{scale\_pos\_weight}: Adjusts the penalty for misclassifying the minority class, typically set based on the ratio of majority to minority class samples to address class imbalance.
\end{itemize}
Beyond this point in the text, for clarity, we use the acronyms HFE for heavy-flavor decay electrons and LFE for light-flavor decay electrons. Similarly, we use HFM and LFM for heavy-flavor and light-flavor decay muons, respectively.

\subsection{Quality Assurance}
In this section, we perform a quality assurance check to ensure the reliability of the input parameter space and the performance of the machine learning models. We plot the correlation matrix in Fig.~\ref{fig_Correlation_matrix} to better understand the interplay between the input parameters. It showcases the Pearson correlation between the pair of two input parameters. A strong correlation indicates a redundancy in the parameter space, which in turn may affect the model's performance. In Fig.~\ref{fig_Correlation_matrix}, the left panel shows the correlation among the input parameters for electrons. We observe no strong correlation among the input parameters. In the right panel of Fig.~\ref{fig_Correlation_matrix}, we plot the correlation in the parameter space for the muons. Here, we observe a negative correlation between \DCAXY~and \DCAZ. This negative correlation exists for muons and is absent for electrons due to the rapidity window of the muons. We consider the muons detected in the forward rapidity window of $2.5 < y < 4$, which in turn generates a correlation between \DCAXY~and \DCAZ. Following this, we plot the importance score of the parameter space in Fig.~\ref{imp_score}. We observe that for electrons, the model has relied mostly on \DCAXY; however, for muons, both \DCAXY~and \DCAZ contribute towards the model training. However, the pseudo-rapidity dependence is very small for both electrons and muons. 

In Fig.~\ref{fig_CM_electron}, we show the confusion matrix corresponding to the two different target classes. Along the X-axis and Y-axis, we have the predicted and true values for heavy and light flavor decay electrons, respectively. We observe an accuracy up to 98\% for HFE as well as LFE. However, the absolute number of true HFE and LFE sheds light on the class imbalance among them. There is a misprediction of around 2\% for both classes. This mainly stems from the fact that we do not consider decay of quarkonia into the HFE category. Finally, in the case of muons in Fig.~\ref{fig_CM_muon}, an accuracy of around 100\% is observed for HFM, meanwhile the accuracy drops to around 91\% for LFM. One can notice a distinct difference in the absolute number of HFM and LFM. This emerges as a result of treating $\pi^{\pm}$ and $K^{\pm}$, as stable particles in PYTHIA8. Thus, preventing them from decaying into $\mu^{\pm}$ and their respective neutrino. Hence, a very small statistics leads to a weakly trained model, resulting in a comparatively lower accuracy.

\section{Results}
\label{sec: Results}
\subsection{Transverse Momentum spectra}
\begin{figure*}
    \centering
    \includegraphics[width = 0.32\textwidth]{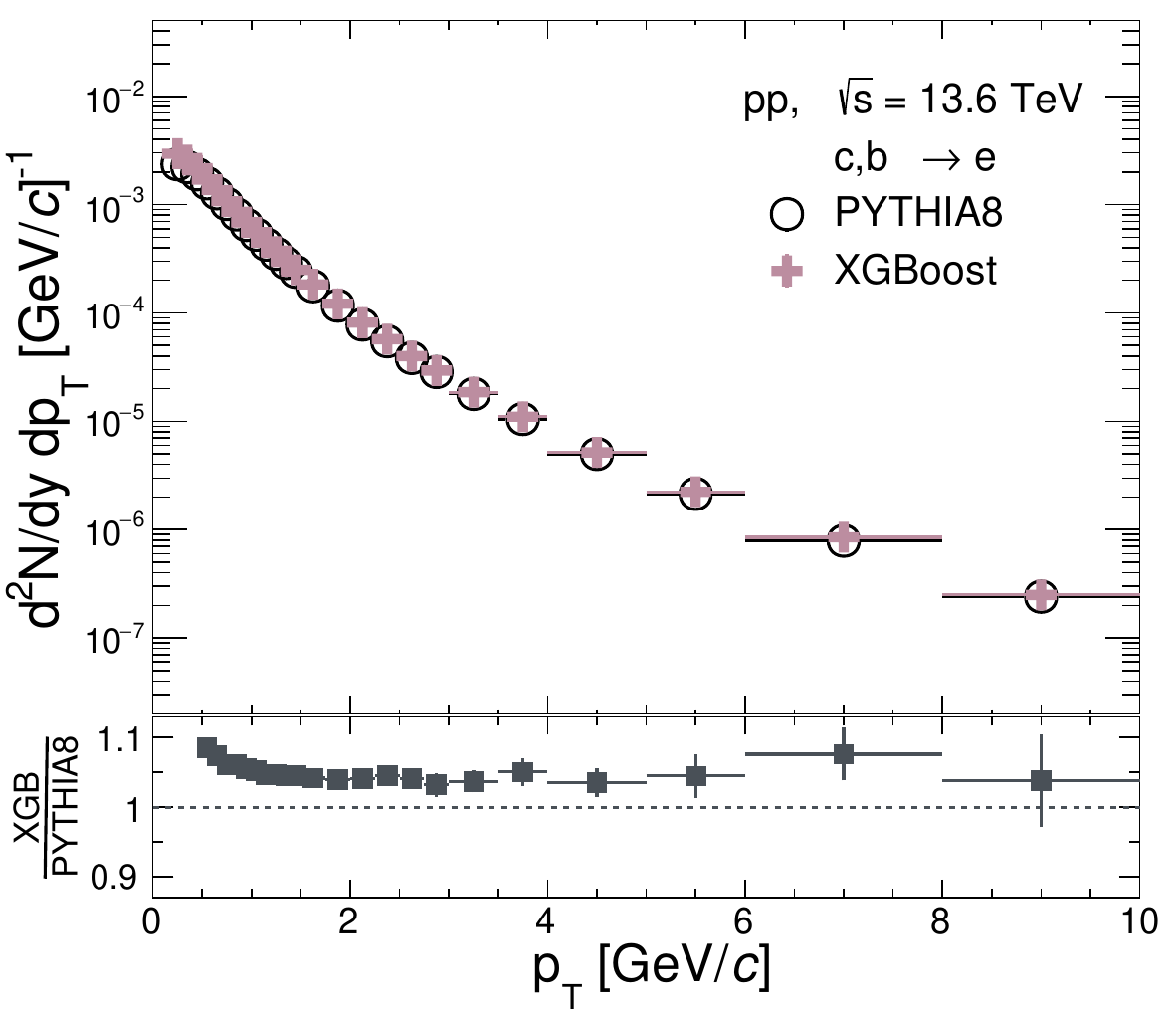} 
    \includegraphics[width = 0.32\textwidth]{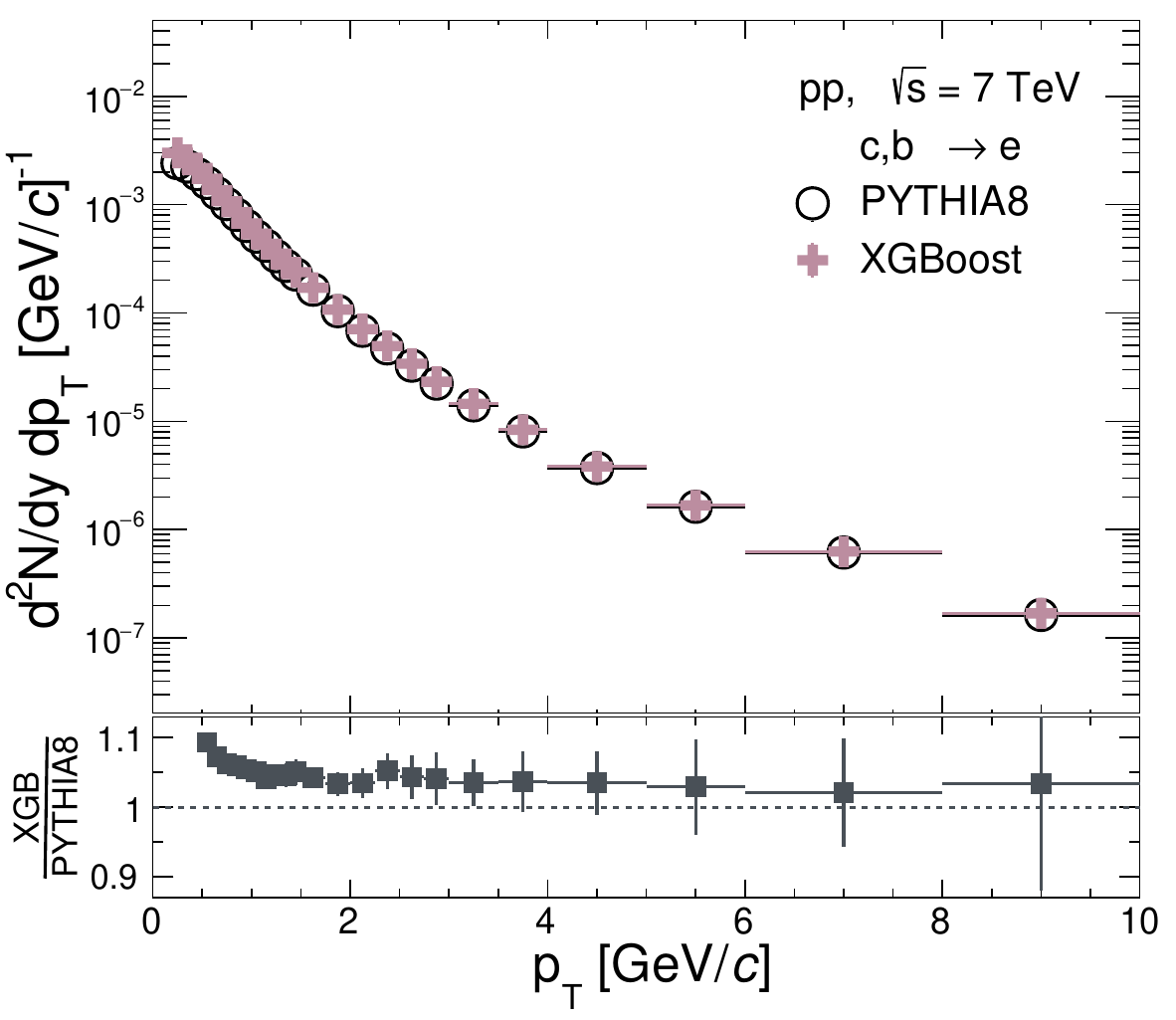} 
    \includegraphics[width = 0.32\textwidth]{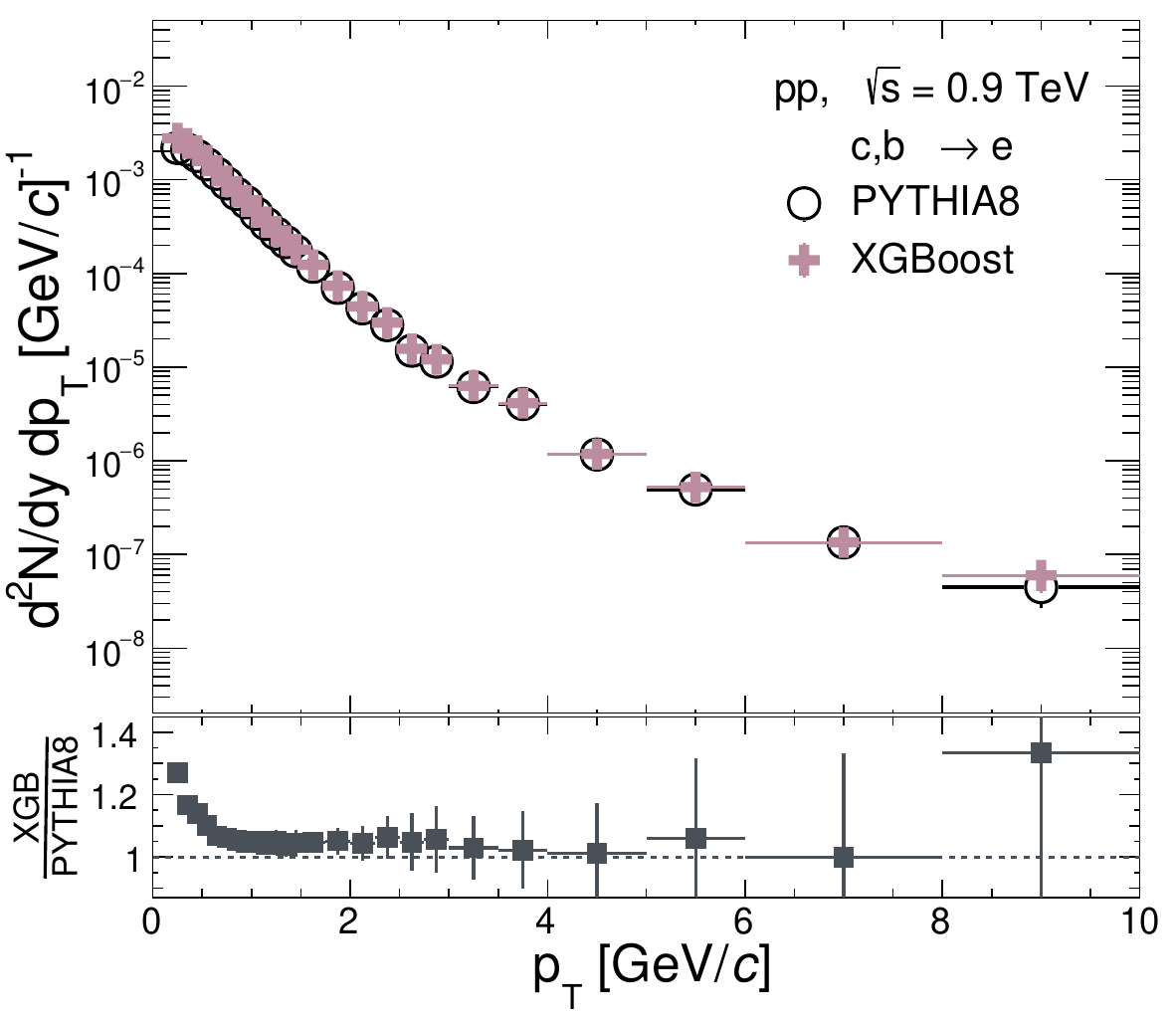} 
    \caption{Transverse momenta spectra of PYTHIA8 generated heavy flavor decay electron in pp collisions in $|\eta|<0.8$ at (a) $\sqrt{s} = 13.6$ TeV (b) $\sqrt{s} = 7$ TeV and (c) $\sqrt{s} = 900$ GeV and predicted by the XGB model.}
    \label{fig_pT_electron}
\end{figure*}
\begin{figure*}
    \centering
    \includegraphics[width = 0.32\textwidth]{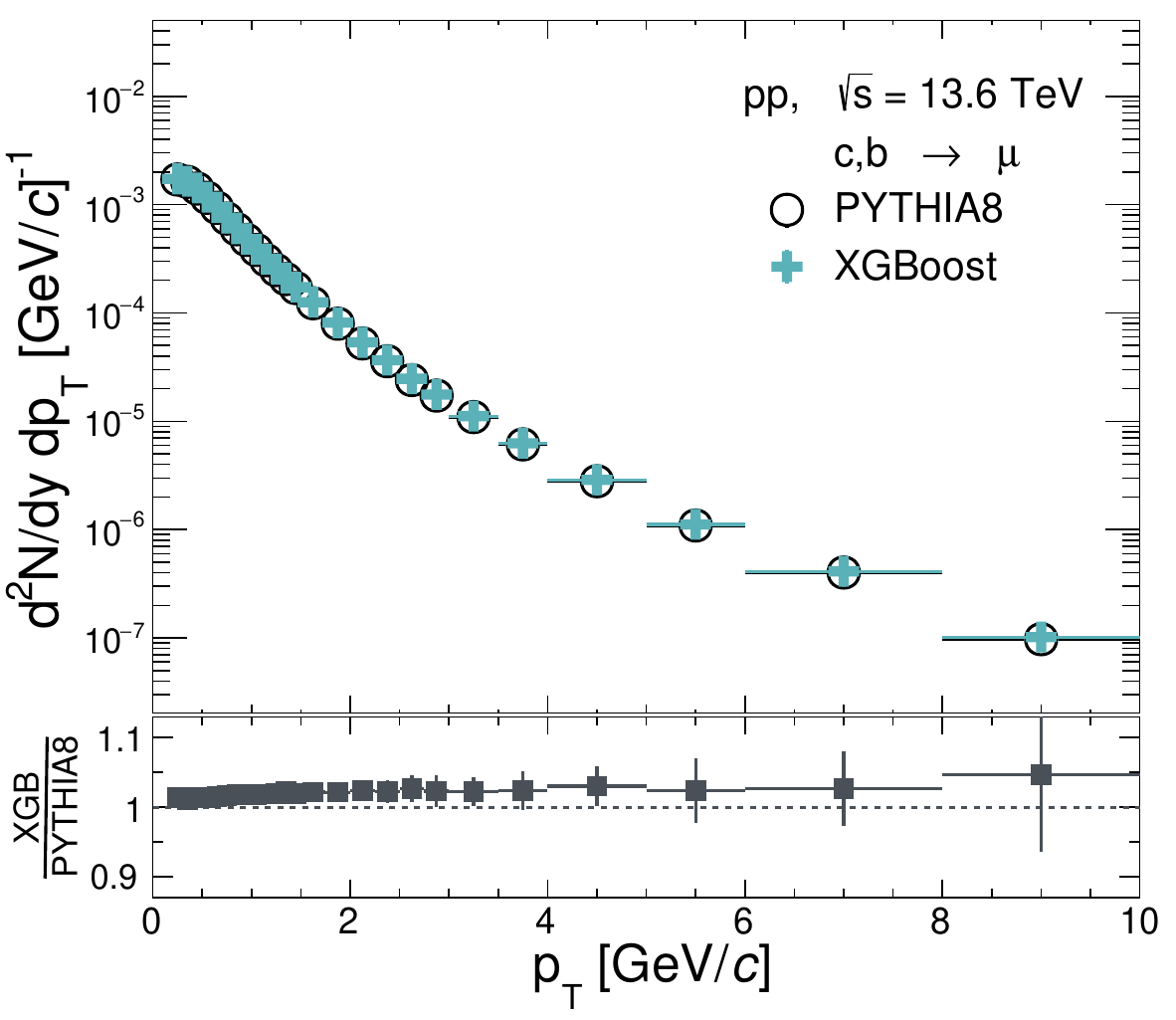} 
    \includegraphics[width = 0.32\textwidth]{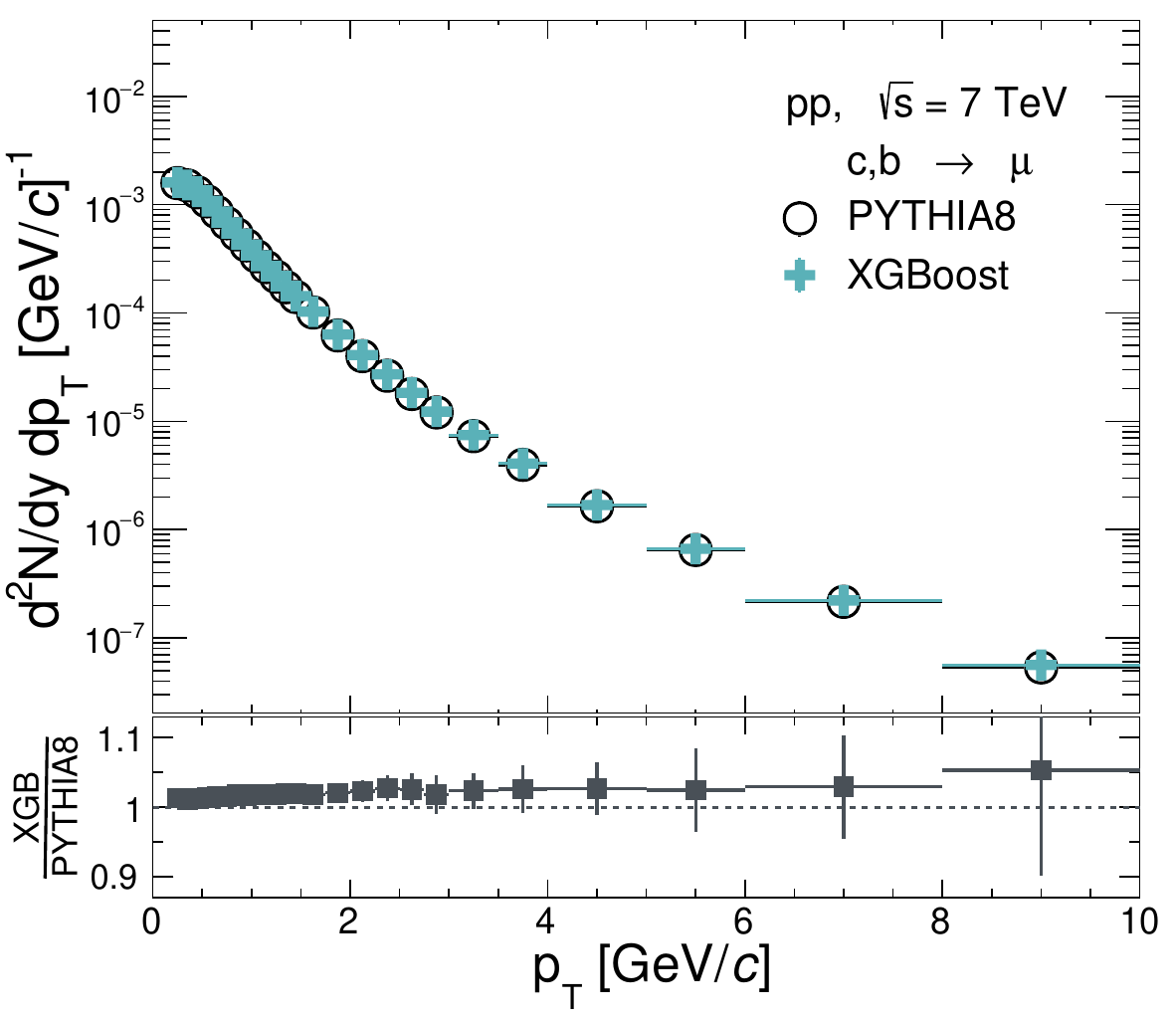} 
     \includegraphics[width = 0.32\textwidth]{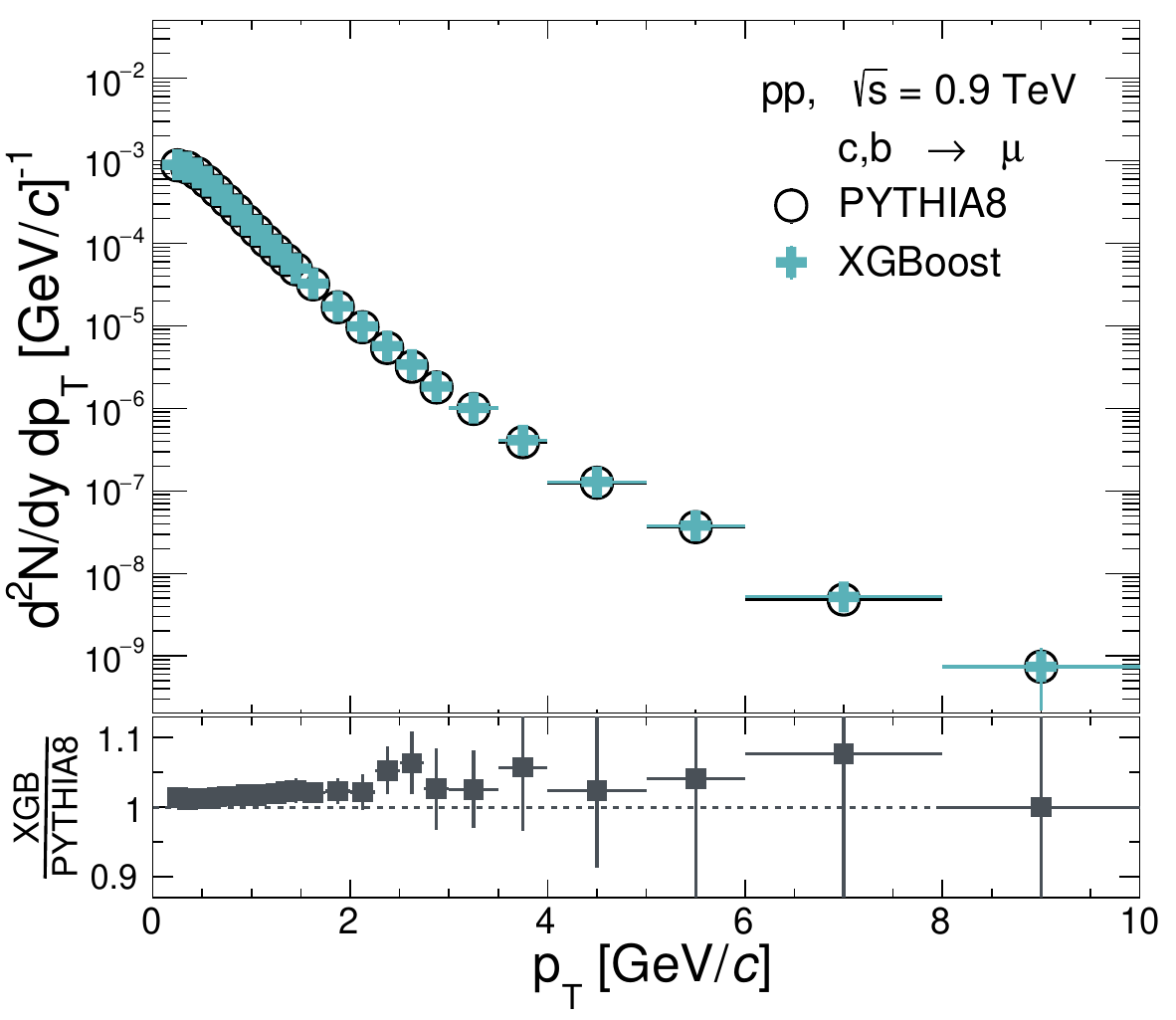} 
    \caption{$p_T$ distribution of PYTHIA8 generated heavy flavor decay muons in pp collisions in $2.5<\eta<4$ at (a) $\sqrt{s} = 13.6$ TeV (b) $\sqrt{s} = 7$ TeV and (c) $\sqrt{s} = 900$ GeV and predicted by the XGB model.}
    \label{fig_pT_muon}
\end{figure*}

In Fig.~\ref{fig_pT_electron}, we plot the transverse momentum~($p_{\text{T}}$) spectra of heavy flavor decay electrons in the mid-rapidity $y<|0.8|$ for three different center-of-mass energies, i.e., $\sqrt{s} = $ 13.6, 7, and 0.9 TeV. The plots include results from PYTHIA8 and the corresponding predictions from the XGBoost model. To have a better understanding of the model's predictions, the ratio of XGBoost to PYTHIA8 yield is shown in the lower panel for each center-of-mass energy. For the low $p_{\text{T}}$ region, the XGBoost predictions deviate from PYTHIA8 data by approximately 10\%, and it sharply decreases to around 5\% for the rest of the transverse momentum bins. This trend is consistently observed across all collision energies. The inaccuracy at low $p_{\text{T}}$ arises from severe class imbalance among the electrons coming from light and heavy flavor. It is important to note that the XGBoost model is trained with PYTHIA8 data for pp collisions at $\sqrt{s} = 13.6 \text{TeV}$. However, the machine learning model performs well in predicting the $p_{\text{T}}$ differential yield across different collision energies, indicating that the model successfully learns and predicts the underlying energy dependence.

Similarly, the heavy flavor hadrons weakly decay into a muon and its respective neutrino. To explore this phenomenon, we plot the $p_{\text{T}}$ differential heavy flavor decay muon yield in Fig.~\ref{fig_pT_muon} in the forward region ($-4.0<y<-2.5$), for three different center-of-mass energies. In this case, the XGBoost to PYTHIA8 yield ratio shows that the models perform really well throughout the entire $p_{\text{T}}$ range for 13.6 and 7 TeV. However, we observe that due to low statistics in the case of $\sqrt{s} = 0.9~\text{TeV}$, we get large statistical error bars as well as slight disagreement with XGBoost and PYTHIA8 data in the intermediate $p_{\text{T}}$. Moreover, the robustness of the model is further tested by predicting data generated with the \texttt{CR:off} tune. The corresponding results are plotted and discussed in Appendix~\ref{App:A}.

\subsection{Self-Normalized Yield of electrons and muons}
\begin{figure*}
    \centering
    \includegraphics[width = 0.45\textwidth]{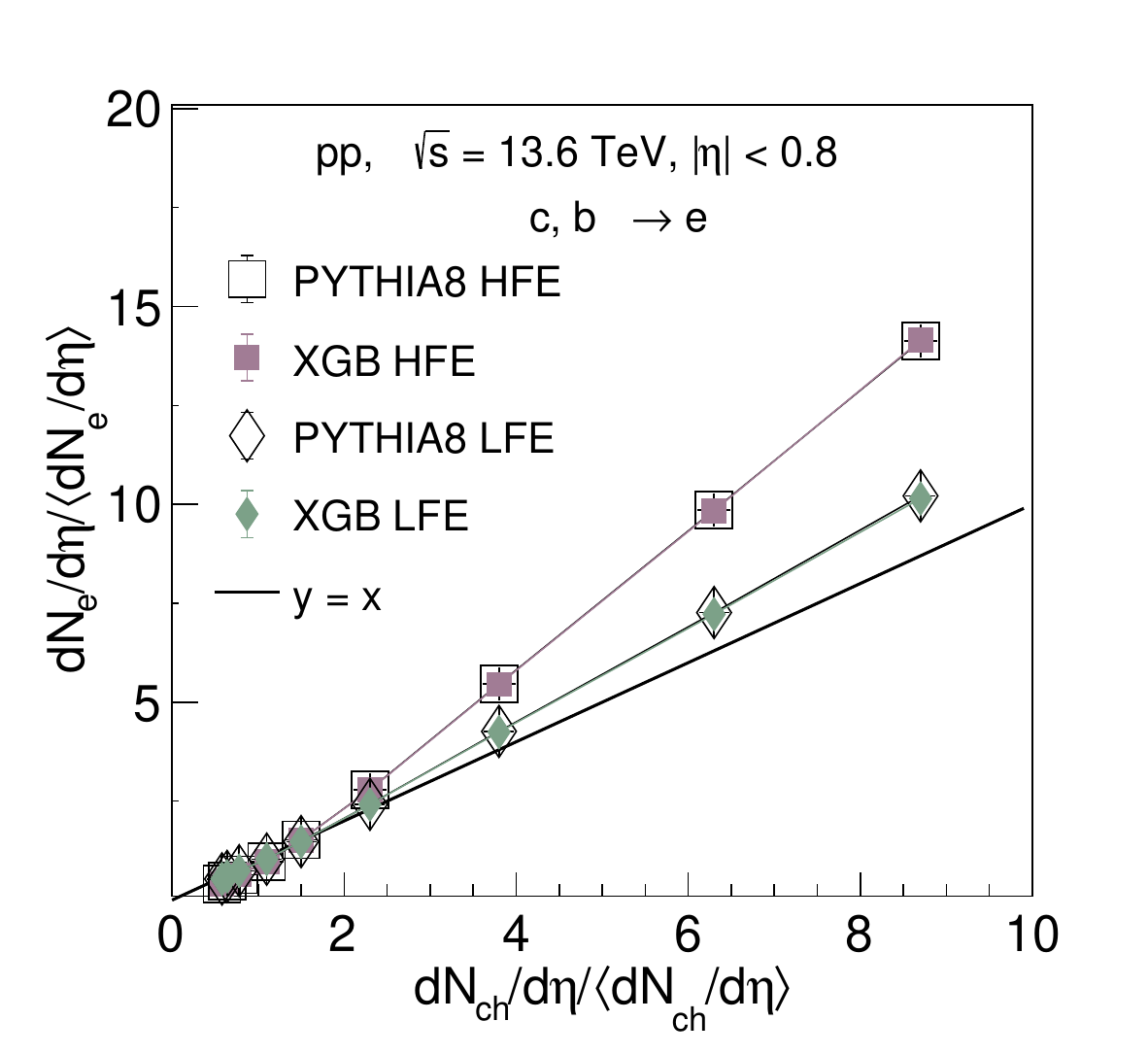} 
    \includegraphics[width = 0.45\textwidth]{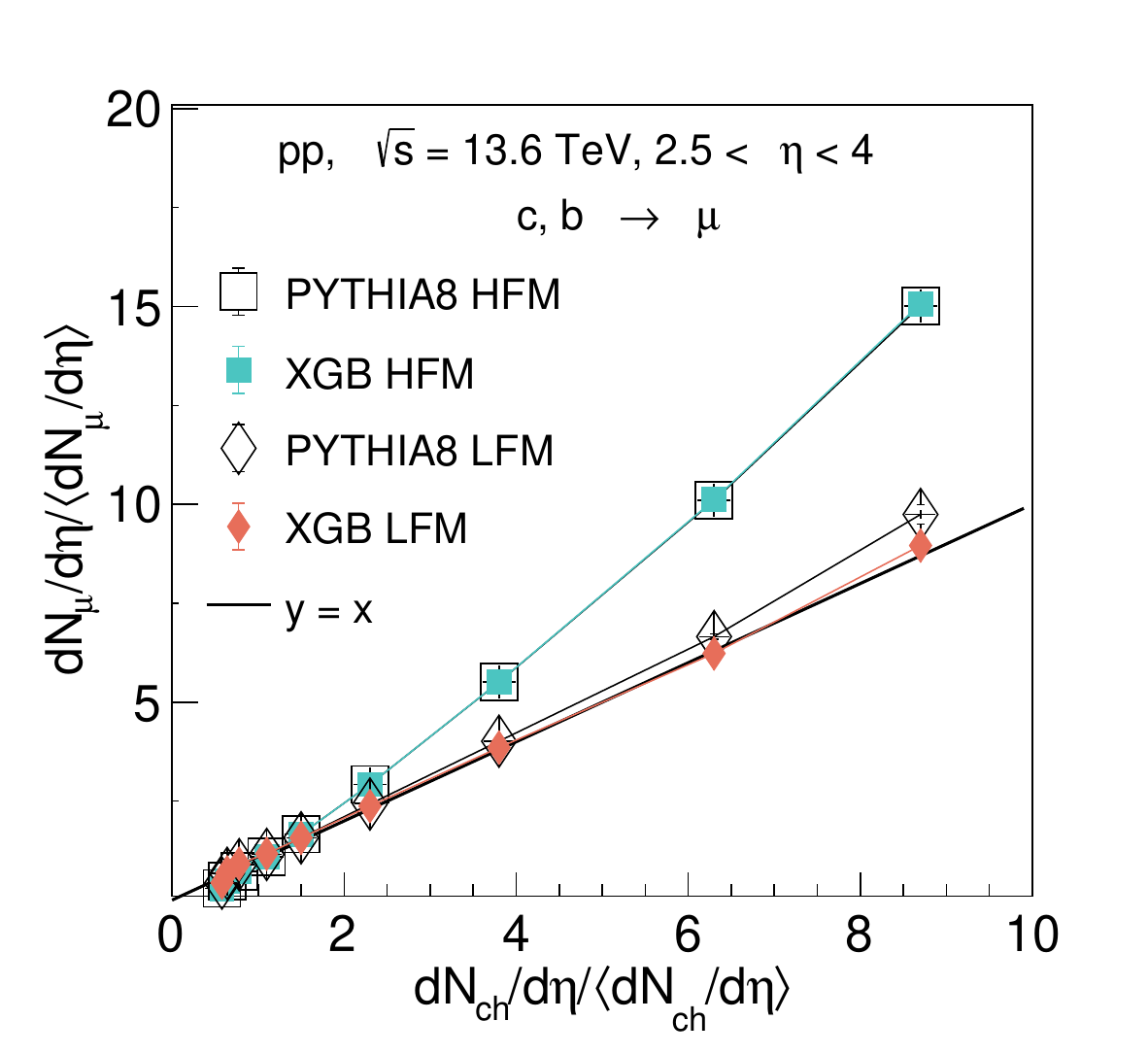} 
    \caption{Self-normalized yield of heavy flavor and light flavor decay electrons (left) and muons (right) as a function of self-normalised charged particle multiplicity density ($dN_{\rm ch}/d\eta/\langle dN_{\rm ch}/d\eta\rangle$) measured at $|\eta|<0.8$ for event selection based on particle multiplicity in forward rapdity, i.e., $-3.7<\eta<-1.7$ and $2.8<\eta<5.1$ in pp collisions at $\sqrt{s}$ = 13.6 TeV estimated using data generated by PYTHIA8 and predicted by the XGB model.}
    \label{fig_SNY}
\end{figure*}
In Fig.~\ref{fig_SNY} (left), we plot the self-normalized yield of electrons in $pp$ collisions at $\sqrt{s} = 13.6~\text{TeV}$, coming from light and heavy flavor decay in the mid-rapidity region, as a function of normalized charged particle multiplicity. To get rid of any autocorrelation bias, the charged particle multiplicity measurement is done in the ALICE V0 region with pseudo-rapidity intervals of $2.8 < \eta < 5.1$ and $-3.7 < \eta < -1.7$. We observe a rising trend for both light and heavy flavor decay electrons, respectively. However, as one would expect, the light flavor decay electrons rise almost linearly, following the $y=x$ curve closely up to a normalized charge particle multiplicity of about 6. However, a non-linear trend can be seen for the heavy flavor decay electrons as it rises very sharply at high multiplicity events. This is because the heavy flavor hadrons are produced along with a jet in the opposite direction, which fragments to increase particle multiplicity in the final state. Therefore, the presence of heavy-flavor hadrons in the final state usually means the charged particle multiplicity of the event is large, or conversely, if the particle multiplicity is large, the probability of production of heavy-flavor hadrons is larger. 


Similarly, in Fig.~\ref{fig_SNY} (right), we study the self-normalized yield of light and heavy flavor decay muons in the forward rapidity, with normalized charged particle multiplicity. One can readily observe that the light-flavor decay muons are increasing linearly as a function of charged particle multiplicity. However, the predictions from the XGBoost model underestimate the PYTHIA8 results slightly. This is evident from the confusion matrix in Fig.~\ref{fig_CM_muon}, the machine learning models fail slightly to understand the production dynamics of light-flavor decay muons. Furthermore, the heavy-flavor decay muons show a stronger-than-linear rise, one similar to the electrons, as a function of multiplicity.

\subsection{$p_{\rm T}$-differential yield ratios}

\begin{figure*}
    \centering
    \includegraphics[width = 0.45\textwidth]{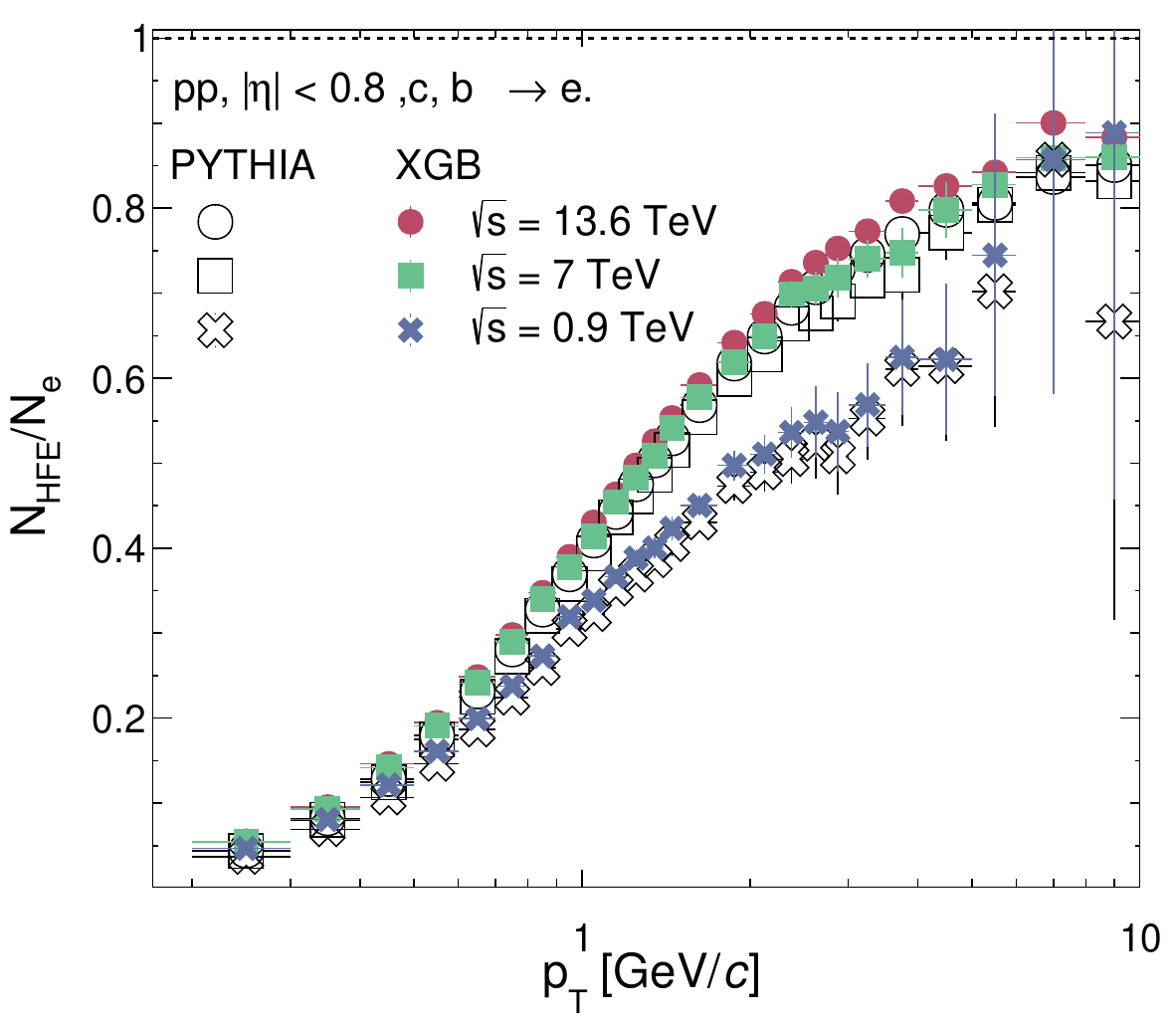} 
    \includegraphics[width = 0.45\textwidth]{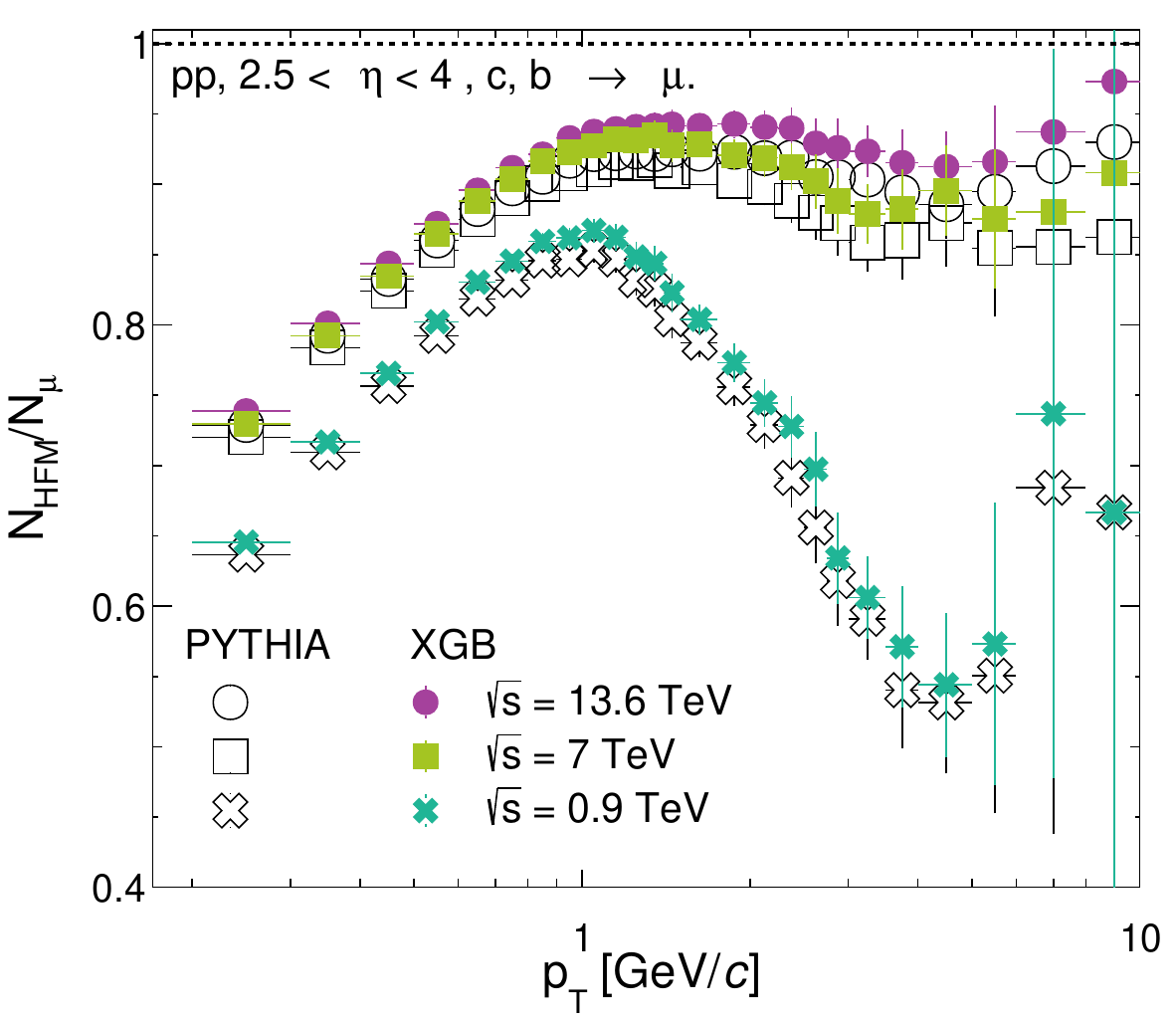} 
    \caption{Ratio of heavy flavor decay electron to total electron (left) and heavy flavor decay muon to total muon (right) at three centre-of-mass energies}
    \label{fig_HFE_Total_ratio}
\end{figure*}
In Fig.~\ref{fig_HFE_Total_ratio}~(left panel), we plot the ratio of heavy flavor decay electrons to the total electron sample, containing contributions from light and heavy flavor decay electrons. We notice a similar trend across all centre-of-mass energies, with an increase in the transverse momentum, the relative yield of the heavy flavor decay electrons increases. At the high $p_{\text{T}}$ region, it tends to reach unity, with a considerable error bar for $\sqrt{s} = 900~\rm{GeV}$. We observe an energy dependence in the intermediate transverse momenta region. The HFE in higher centre-of-mass energy is more abundantly produced in the intermediate $p_{\text{T}}$ region. This increase in the ratio with $p_{\text{T}}$ reflects the fact that electrons from light-flavor decays predominantly populate the low-$p_{\text{T}}$ region, whereas heavy-flavor decay electrons dominate at higher $p_{\text{T}}$ due to their harder $p_{\text{T}}$ spectra resulting from the heavy quark mass and their production mechanism. The observed energy dependence in the intermediate $p_{\text{T}}$ region may be attributed to the enhanced phase space available for heavy-flavor production at higher centre-of-mass energies, leading to a relative increase in their contribution compared to electrons coming from light flavor. A similar trend can be seen for the fraction of $J/\psi$ produced from b-hadron decay, as reported in~\cite{Prasad:2023zdd, ALICE:2021edd}. These results indicate that regardless of the particle studied, the contribution from heavier particles becomes prominent at higher $p_{\rm{T}}$. 
In the right panel of Fig.~\ref{fig_HFE_Total_ratio}, we plot the fraction of heavy flavor decay muons to total muons. We observe a similar increase in the ratio in the low $p_{\rm{T}}$ region up to 1 GeV/\textit{c}. However, for $p_{\rm{T}}>1~\rm{GeV/\textit{c}}$, we observe a dip in the ratio. This dip is very prominent at low center-of-mass energy, and for $\sqrt{s}=13$ and $7$ TeV, the dip in the ratio is very small. One possible reason for this significant dip at 0.9 TeV is due to limited phase space for producing a heavy flavor hadron, which in turn will decay to high $p_{\rm{T}}$ muons, in the forward rapidity region. Regardless of this, we observe a good agreement between the XGB predictions and PYTHIA data.  However, a slight disagreement is observed at high $p_{\rm{T}}$ within large error bars.

\subsection{Lepton-hardon azimuthal correlations}

\begin{figure}
    \centering
    \includegraphics[width = 0.5\textwidth]{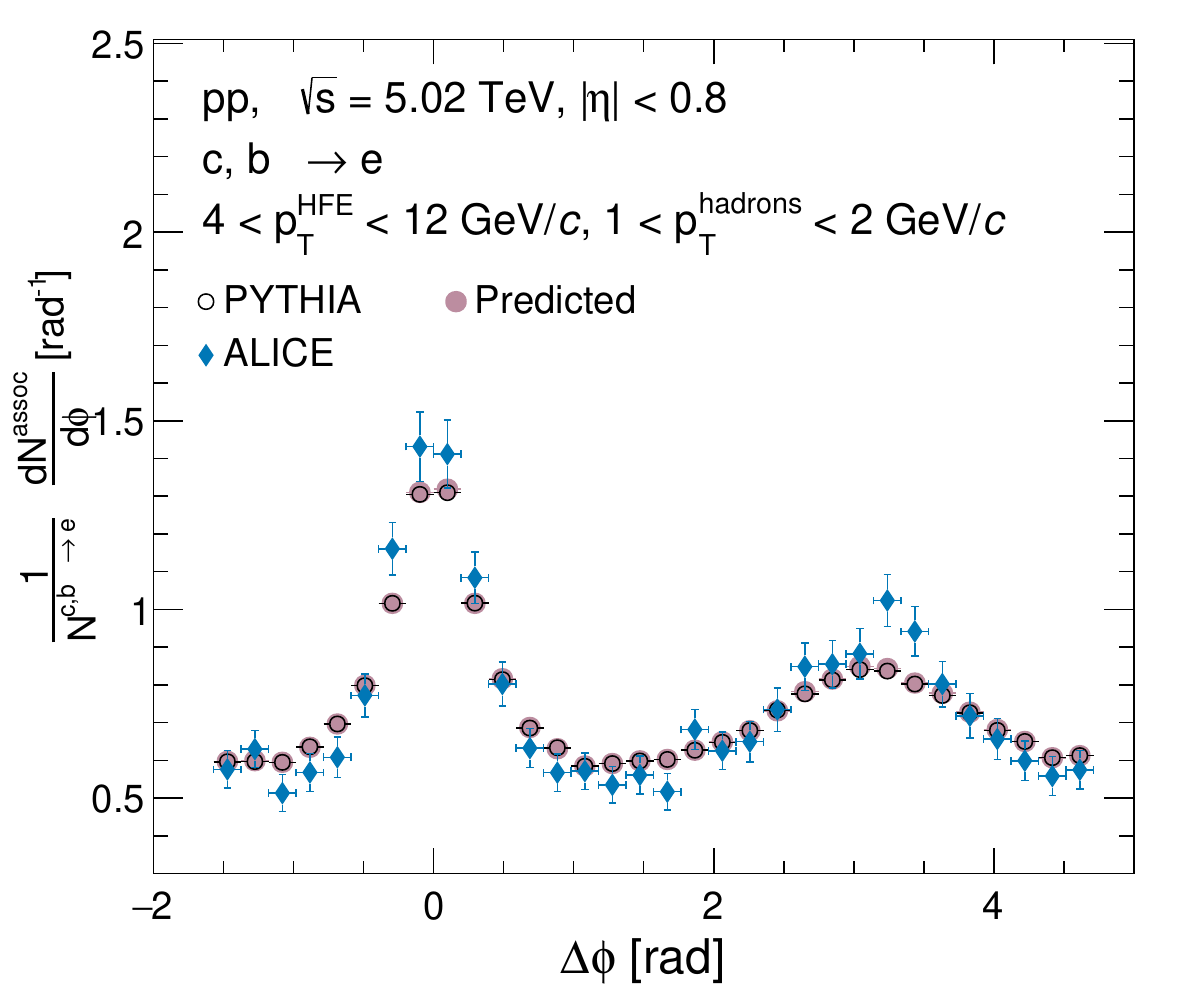} 
    \caption{Comparison of azimuthal correlation distribution of HFE at $4<p_{\rm T}^{\rm HFE}<12$ GeV/\textit{c} and charged hadrons at $1<p_{\rm T}^{\rm hadrons}<2$ GeV/\textit{c} in pp collisions $\sqrt{s}$ = 5.02 TeV between data generated by PYTHIA8, ALICE data~\cite{ALICE:2023kjg} and predicted by the XGB model. A scaling factor of 0.5 is used to get an intuitive comparison between ALICE data and our work.}
    \label{fig_azimuthal_correlation}
\end{figure}
In Fig.~\ref{fig_azimuthal_correlation}, we plot the azimuthal correlation among heavy-flavor decay electrons and hadrons, with heavy-flavor decay electrons $p_{\text{T}}$ in the range of $4$--$12~\text{GeV/\textit{c}}$. We consider all the hadrons with transverse momenta within $1$--$2~\text{GeV/\textit{c}}$ and compare our results with ALICE data for $pp$ collision at $\sqrt{s}=5.02~\text{TeV}$. We find that the prediction from XGBoost aligns well with the PYTHIA results. Moreover, we observe a remarkable agreement of our results with ALICE data. The near-side peak around $\Delta\phi = 0$ is primarily due to the fragmentation of the same parton that produced the heavy-flavor decay electron, whereas the away-side peak around $\Delta\phi \sim \pi$ originates from the fragmentation of the recoiling parton. The shape and relative strength of these peaks are sensitive to the production mechanism and hadronization process of heavy quarks. The consistency of the correlation pattern between the XGBoost-predicted electrons and those from PYTHIA and ALICE indicates that the classifier preserves the intrinsic angular correlation features of heavy-flavor decay electrons, without introducing any biases in the angular correlation.

\section{Summary}
\label{sec: Summary}
In this study, we segregate heavy-flavor decay electrons and muons directly from track-level information using machine learning algorithms. This method holds the potential to replace more traditional methods like cocktail fitting. We performed a binary classification of heavy-flavor decay leptons and light-flavor decay leptons using the XGBoost machine learning algorithm. The input features are constructed from topological and kinematic variables sensitive to the weak decay of heavy-flavor hadrons into electrons and their respective neutrinos. This includes the distance of closest approach in the transverse plane as well as along the beam direction. We train the models \pT~bin-wise, separately for electrons, in the mid rapidity, and for muons, in the forward rapidity. We observe that the model successfully distinguishes the electrons coming from the heavy-flavor and light-flavor sectors, with an accuracy of 98\%. However, for muons, we observe an unprecedented accuracy of approximately 100\% for heavy-flavor decay muons. On the other hand, due to much smaller statistics of light-flavor decay muons, the model could predict only up to 91\%.

Furthermore, we predict the HFE and HFM yields at different center-of-mass energies, i.e., $\sqrt{s}= 0.9$, $7$, and $13.6$ TeV. We observe that for electrons, the predictions at low \pT~are off by a margin of around 10\%. However, the model is relatively more accurate for mid and high \pT, although for  $\sqrt{s}= 0.9$ TeV, the error bars at high \pT~are significant due to low statistics. Similarly, we plot the transverse momentum spectra for muons. Here, we observe a comparatively more stable prediction throughout the entire \pT~range. Regardless, in a very similar fashion, we observe large error bars for $\sqrt{s}= 0.9$ TeV. In addition, we study the self-normalized yield of HFE, LFE, HFM, and LFM as a function of normalized charged particle multiplicity. For electrons at mid rapidity, we observe a nearly linear rise for LFE, while HFE shows a non-linear rise in high-multiplicity events. The XGBoost model successfully predicts the self-normalized yield across the entire range of charge particle multiplicity. For muons in the forward rapidity region, the LFM shows a similar linear rise, and the HFM exhibits a strong non-linear trend with the charge particle multiplicity. However, we observe a disagreement for the LFM between XGBoost and PYTHIA8, appearing as a consequence of low accuracy for the LFM. Moreover, we investigate the ratio of HFE to total electron yield and HFM to total muon yield as a function of \pT. For electrons, we observe that the ratio increases monotonically with \pT~and approaches unity at high \pT. A clear energy dependence is visible in the intermediate \pT~region. For muons, a similar rise is observed; however, a significant dip in the ratio is observed at $\sqrt{s}=0.9$ TeV. This dip, even though visible at higher energies of 13 and 7 TeV, is not significant. Finally, we study azimuthal correlation between HFE and charged hadrons in $pp$ collisions at $\sqrt{s}=5.02$ TeV. One can observe clear near-side and away-side peaks, consistent with ALICE experimental data. This work can be extended by incorporating realistic detector effects, such as electrons originating from the photon conversion in the detector material. These contributions would provide a more complete and realistic study of the identification of heavy flavor decay leptons in experiments. With the upcoming facilities like FCC, NICA, and EIC, combined with recent developments in machine-learning techniques, the study of heavy flavor decay leptons provides a novel and unique way to study the heavy flavor sector.

\section*{Acknowledgment}
K.G. acknowledges financial support from the Prime Minister's Research Fellowship (PMRF), Government of India. S.P. gratefully acknowledges the financial aid from the University Grants Commission (UGC), Government of India. The authors (K.G., S.P., and R.S.) acknowledge the DAE-DST, Government of India, funding under the mega-science project “Indian participation in the ALICE experiment at CERN” bearing Project No. SR/MF/PS-02/2021-IITI(E-37123).

\newpage
\appendix
\onecolumngrid

\section{Comparison with other PYTHIA8 tunes}
\label{App:A}
To further strengthen our validation strategy, we use the model (trained with \texttt{CR: on, CRMode: 2}) to predict data generated with a different tuning (\texttt{CR: Off}). The specific change in the tune alters the hadronization mechanism thus introduces a change in the secondary vertex positions of the produced particles and significantly affects the particle multiplicity (light and heavy flavour differently) in the final state. The outcomes of this test are briefly described as follows.
\begin{figure}[H]
    \centering
    \includegraphics[width=0.4\linewidth]{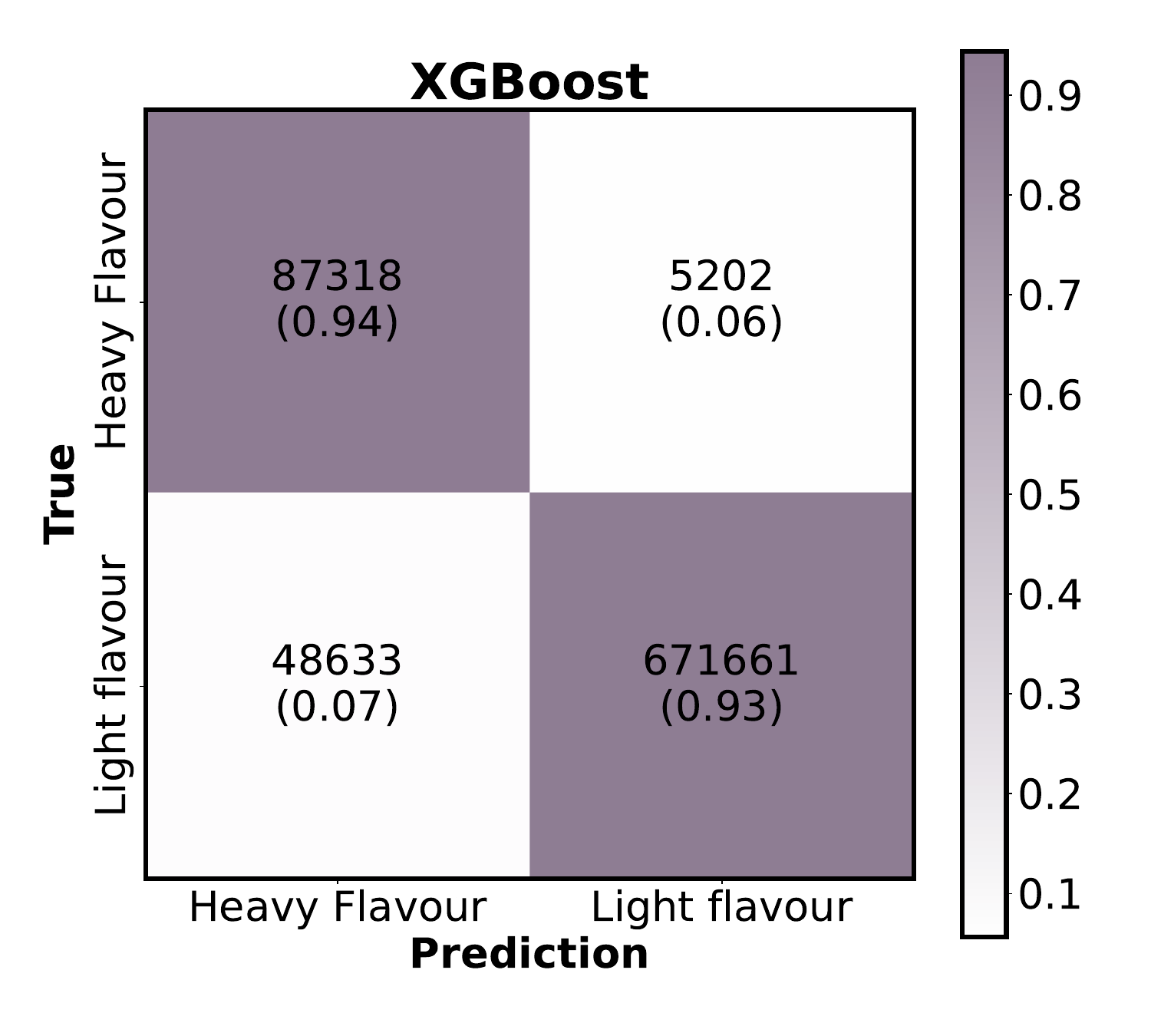}
    \includegraphics[width=0.4\linewidth]{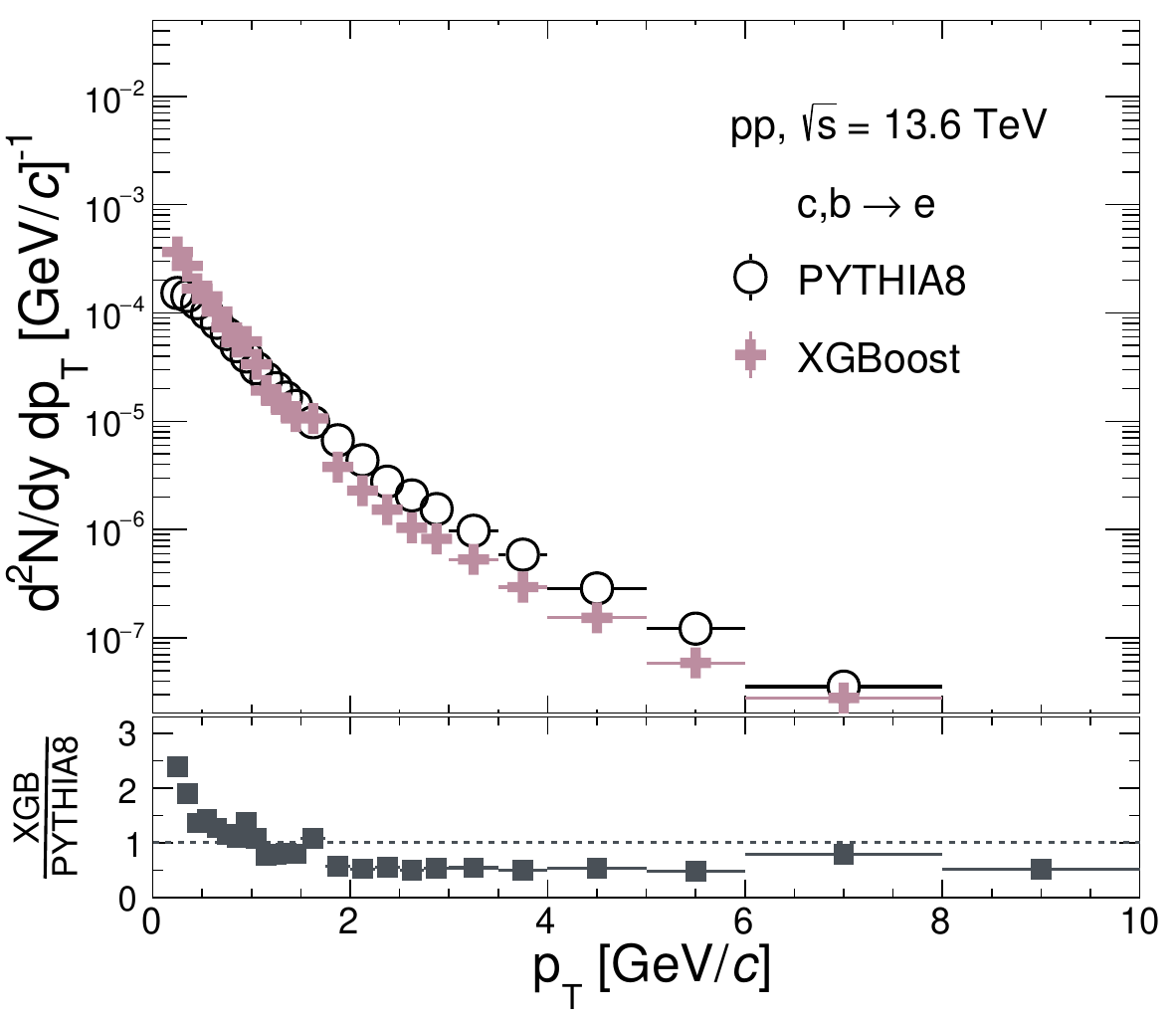}
    \caption{Left: Confusion matrix with CR:Off testing data. Right: Predicted and PYTHIA $p_{\mathrm{T}}$ spectra and their ratio.}
    \label{Fig:CROff_electron}
\end{figure}
In Fig.~\ref{Fig:CROff_electron} (left), we present the confusion matrix obtained using the new testing sample (\texttt{CR: Off}). A slight decrease in performance is observed compared to the accuracy of 98\% reported in Fig.~\ref{fig_CM_electron}: the accuracy for HFE decreases to 94\%, while for LFE it decreases to 93\%. This indicates that the model captures the physics-driven correlations associated with the \texttt{CR:On} configuration. At the same time, achieving an accuracy above 90\% when evaluated on a sample generated with a different tuning demonstrates a certain degree of robustness of the classifier.
\begin{figure}[H]
    \centering
    \includegraphics[width=0.4\linewidth]{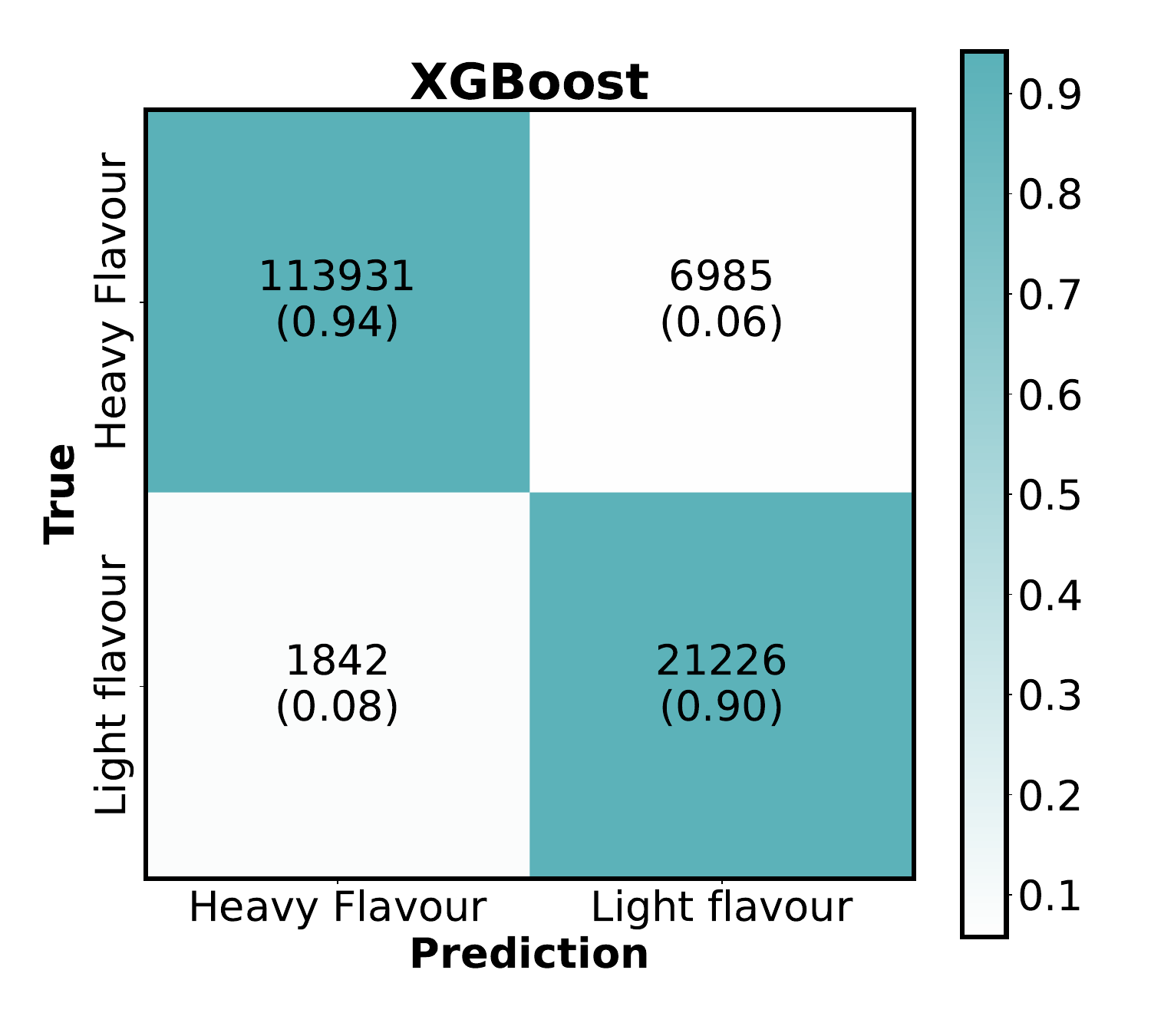}
    \includegraphics[width=0.4\linewidth]{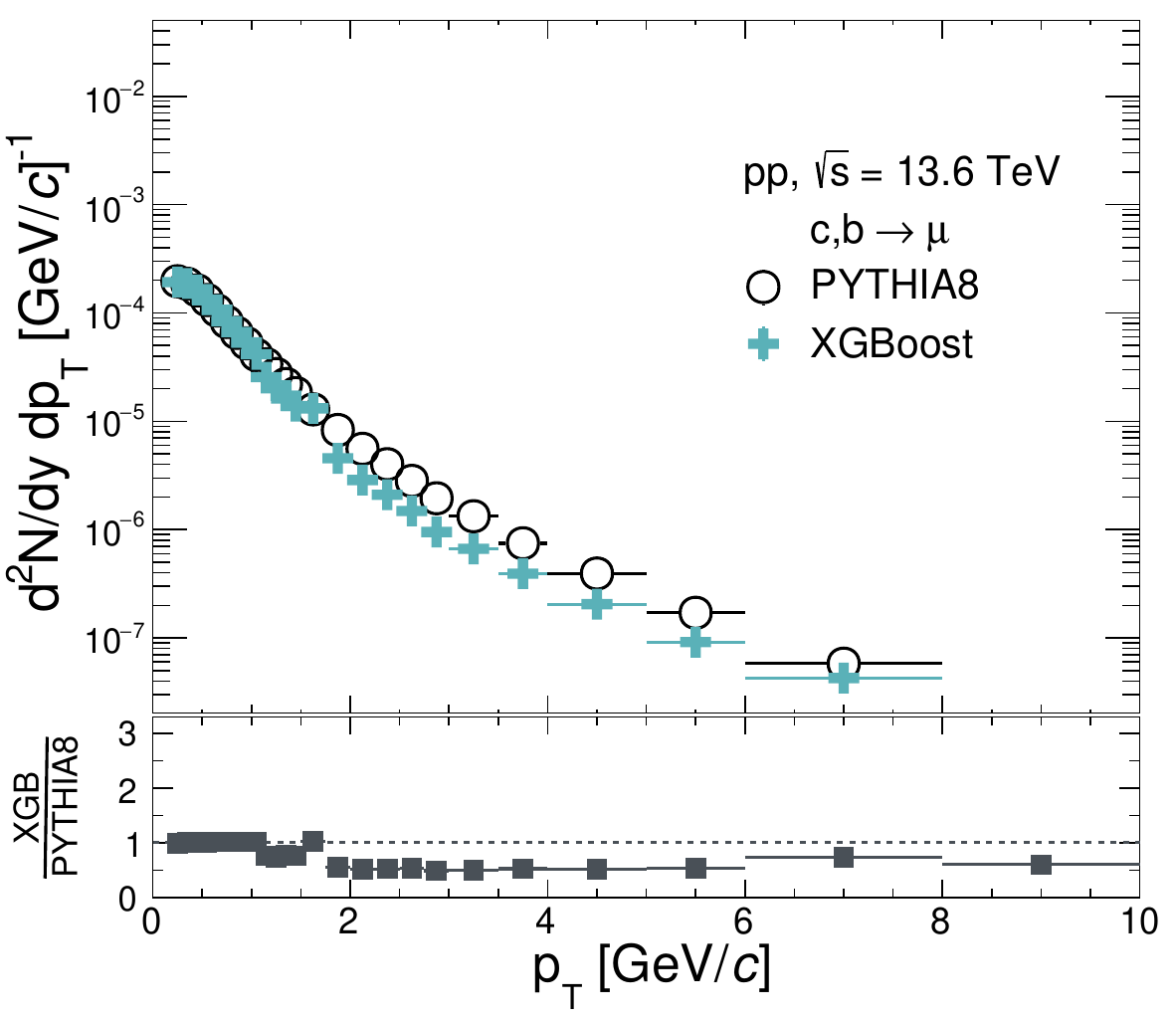}
    \caption{Left: Confusion matrix with CR:Off testing data. Right: Predicted and PYTHIA $p_{\mathrm{T}}$ spectra and their ratio.}
    \label{Fig:CROff_muon}
\end{figure}
In addition, the $p_{\mathrm{T}}$ spectra are shown in the right panel of Fig.~\ref{Fig:CROff_electron} to further quantify the deviation from the PYTHIA prediction. A larger discrepancy is observed in the low-$p_{\mathrm{T}}$ region, which can be attributed to the class imbalance discussed in Sec.~\ref{sec:training}.
We perform the same test for the heavy-flavor decay muons. In Fig.~\ref{Fig:CROff_muon}, we plot the confusion matrix (left) and $p_{\mathrm{T}}$ spectra (right) for the heavy-flavor decay muons. We observe an accuracy of around 94\% for light-flavor decayed muons and 90\% for heavy-flavor decayed muons. Moreover, the predicted $p_{\mathrm{T}}$ spectra shows a reasonable qualitative agreement with PYTHIA results throughout the complete $p_{\mathrm{T}}$ range.

\section{Distribution of DCA}
\label{App:B}
\begin{figure}[H]
    \centering
    \includegraphics[width=0.4\linewidth]{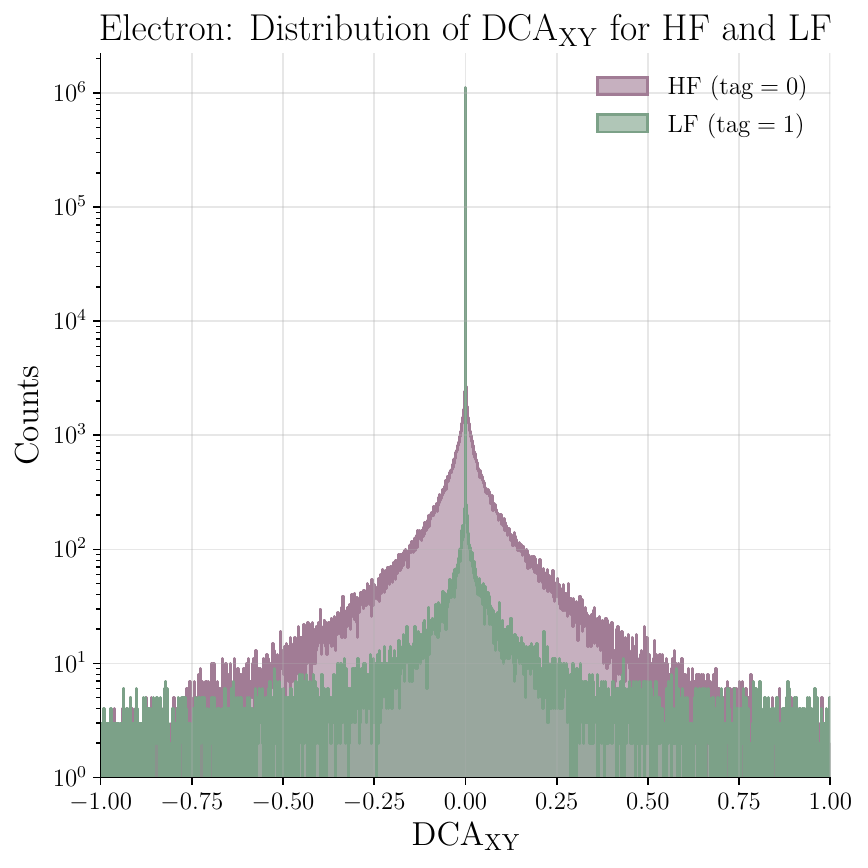}
    \includegraphics[width=0.4\linewidth]{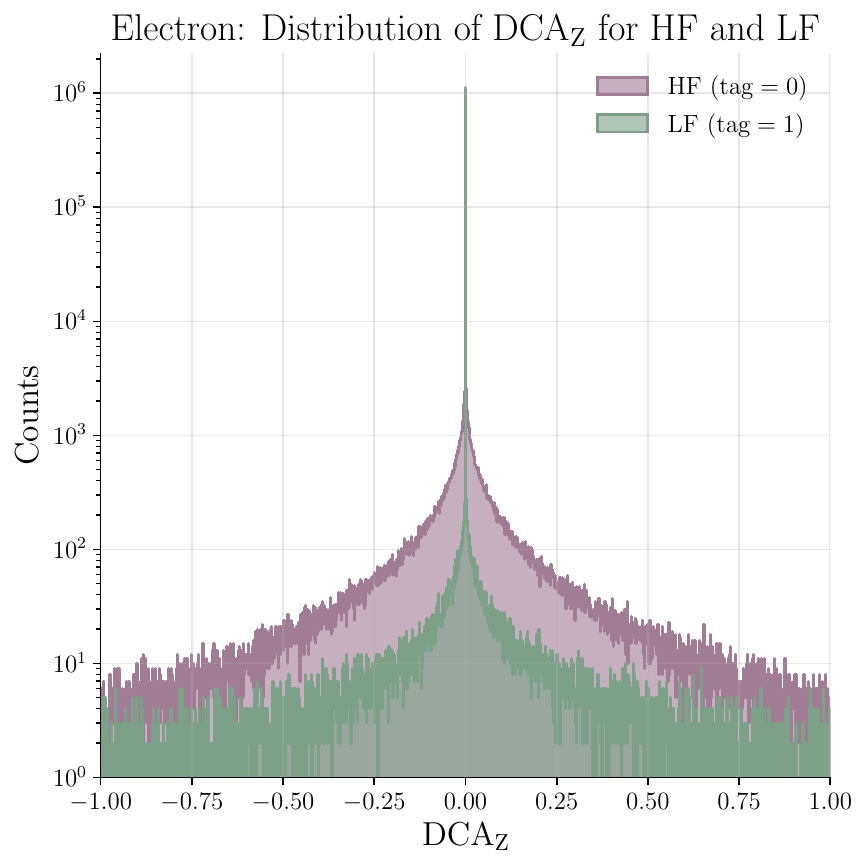}
    \caption{Distributions of the transverse (\DCAXY) and longitudinal (\DCAZ) distance of closest approach for electrons, shown separately for heavy-flavor and light-flavor decay components.}
    \label{fig:DCA_electron}
\end{figure}

\begin{figure}[H]
    \centering
    \includegraphics[width=0.4\linewidth]{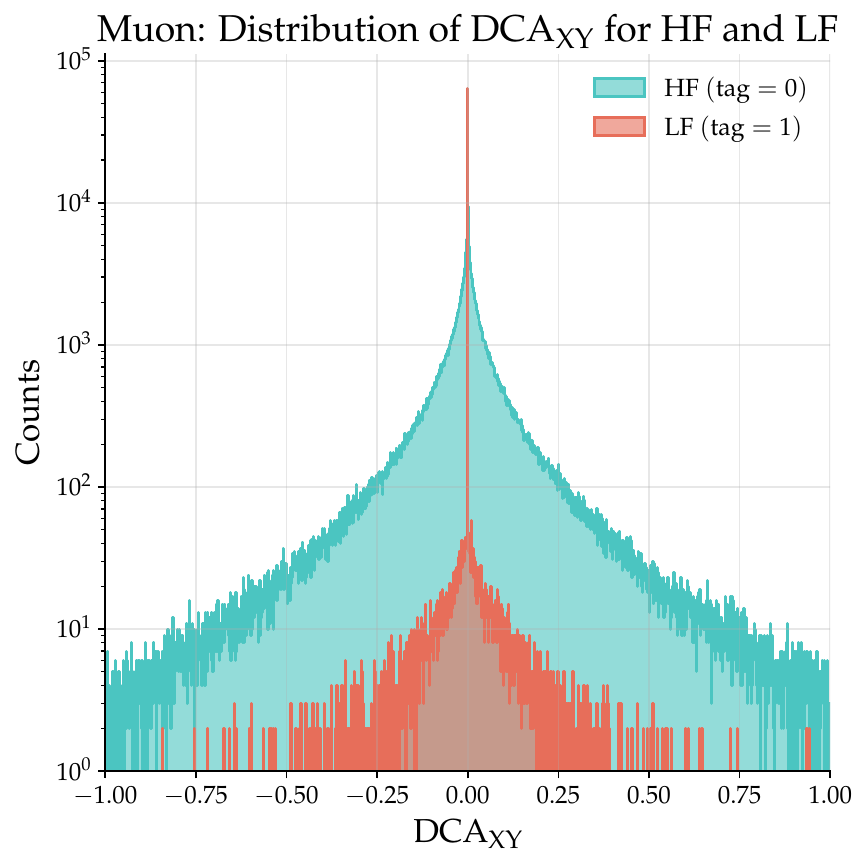}
    \includegraphics[width=0.4\linewidth]{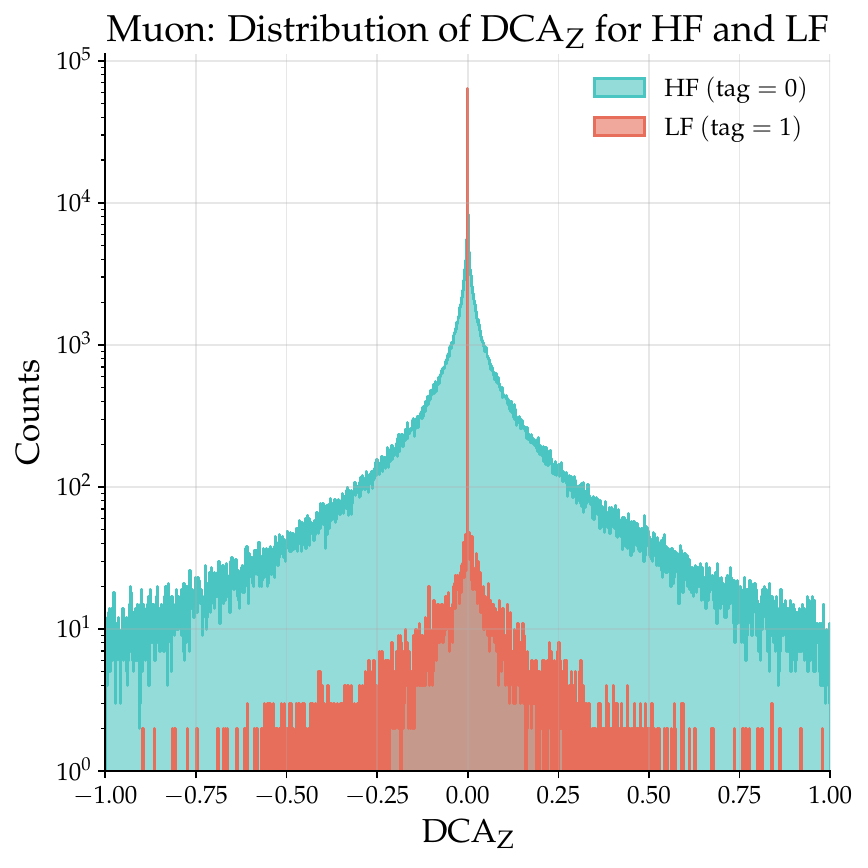}
   \caption{Distributions of the transverse (\DCAXY) and longitudinal (\DCAZ) distance of closest approach for muons, shown separately for heavy-flavor and light-flavor decay components.}
    \label{fig:DCA_muon}
\end{figure}

Figure~\ref{fig:DCA_electron} shows the distributions of the distance of closest approach for electrons, separated into heavy-flavor (HF) and light-flavor (LF) decay components. The left panel presents the transverse component, \DCAXY, while the right panel shows the longitudinal component, \DCAZ. A clear separation between HF and LF electrons is observed in both \DCAXY~and \DCAZ. The LF electrons exhibit a sharply peaked distribution around zero DCA, reflecting their predominantly prompt origin. In contrast, the HF electrons show significantly broader distributions with pronounced tails extending to larger DCA values, arising from the displaced decay vertices of heavy-flavor hadrons. The consistent behavior observed in both transverse and longitudinal directions demonstrates that DCA information provides strong discriminating power for heavy-flavor identification, motivating its use as a key input in machine-learning-based classification of electron sources.
Figure~\ref{fig:DCA_muon} shows the distributions of the distance of closest approach for muons, separated into heavy-flavor (HF) and light-flavor (LF) decay components. Similar to the electron case, a clear distinction between HF and LF muons is observed in both DCA components. The LF muons are characterized by a narrow distribution centered around zero DCA, and the HF muons exhibit significantly broader distributions with extended tails toward larger DCA values, reflecting the finite decay length of heavy-flavor hadrons. The persistence of this separation in both transverse and longitudinal directions indicates that DCA-based observables retain strong sensitivity to heavy-flavor decay topologies in the forward rapidity region. Moreover, the spread in the LF electrons and muons undergoing strong decay is arising from the displacement in the production vertices.

\section{Application of SMOTE}
\begin{figure}[H]
    \centering
    \includegraphics[width=0.4\linewidth]{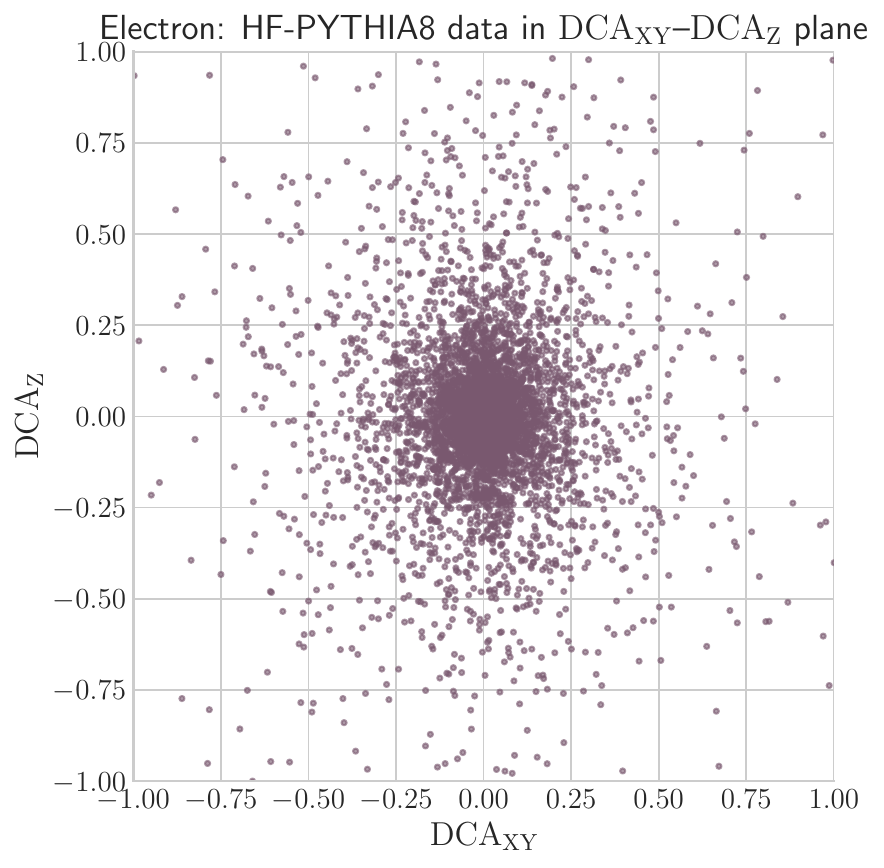}
    \includegraphics[width=0.4\linewidth]{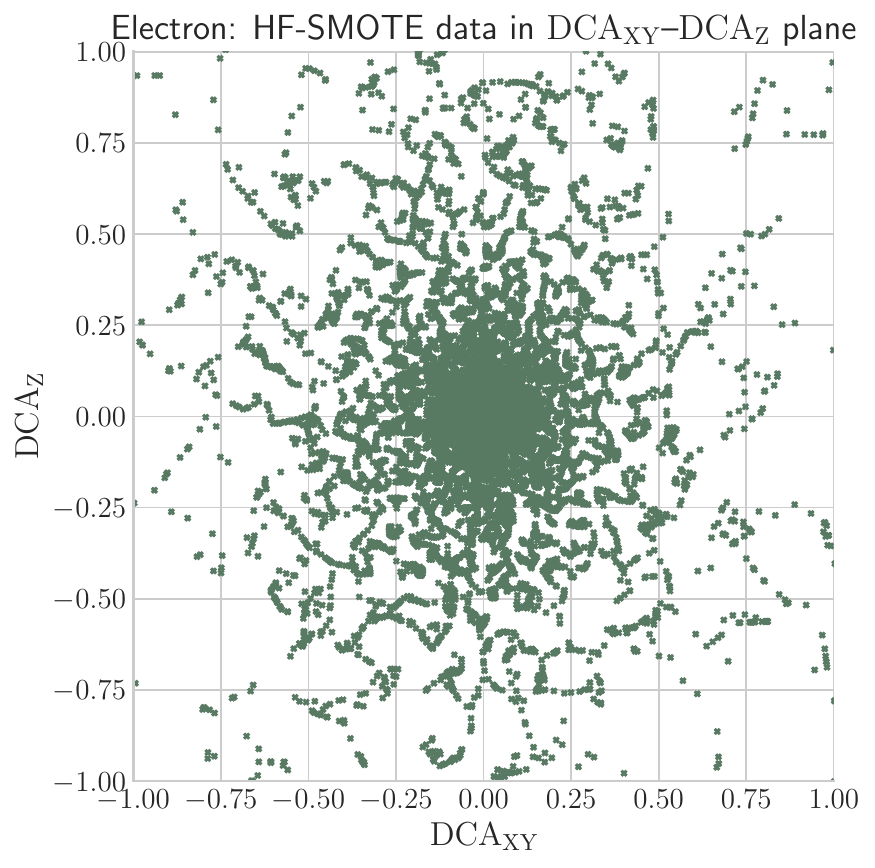}
    \caption{Scatter distributions of heavy-flavor decay electrons in the \DCAXY–\DCAZ~plane for the original dataset (left) and after applying SMOTE (right).}
    \label{fig:scatter-SMOTE}
\end{figure}

Figure~\ref{fig:scatter-SMOTE} illustrates the effect of SMOTE applied to heavy-flavor (HF) decay electrons in the \DCAXY–\DCAZ~plane. In the original dataset, LF electrons dominates, resulting in a significant class imbalance that can introduce bias in machine-learning-based classification, favoring the majority class. Oversampling the minority HF class using SMOTE mitigates this bias by generating synthetic samples that populate the feature space more uniformly, thereby improving the representation of heavy-flavor decay topologies during training.
The left panel shows the scatter distribution of HF decay electrons from the original PYTHIA8 simulation in the \DCAXY–\DCAZ~plane. The distribution is concentrated around the origin, with a limited population in the outer regions of the phase space, reflecting the available HF statistics in the raw sample. The right panel presents the corresponding distribution after applying SMOTE. While preserving the overall structure and correlations between \DCAXY~and \DCAZ, the SMOTE sample exhibits a significantly enhanced coverage of the phase space, particularly in regions sparsely populated in the original dataset. This improved sampling reduces statistical fluctuations and limits the influence of class imbalance, leading to a more stable and unbiased training of the model.

\begin{figure}[H]
    \centering
    \includegraphics[width=0.4\linewidth]{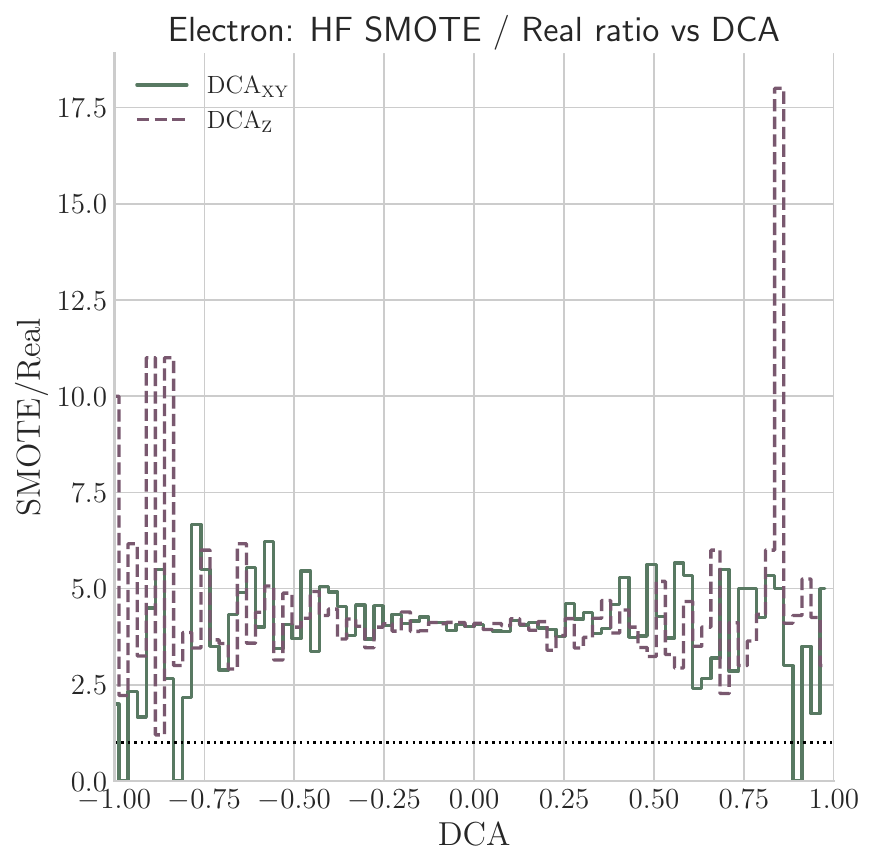}
    \caption{Ratio of SMOTE-generated to real heavy-flavor decay electrons as a function of \DCAXY~and \DCAZ.}
    \label{fig:ratio-SMOTE}
\end{figure}
Figure~\ref{fig:ratio-SMOTE} shows the ratio of SMOTE-generated to real heavy-flavor decay electrons as a function of DCA for both \DCAXY~and \DCAZ. This ratio provides a quantitative measure of how the SMOTE procedure enhances the population of the feature space relative to the original dataset. Over the central DCA region, the ratio remains approximately constant at a value of about four, indicating that the SMOTE sample contains roughly four times more entries than the real data in this region. Larger deviations observed at high absolute DCA values are primarily driven by the limited statistics in the original sample, where SMOTE increases the representation of sparsely populated regions. The similar behavior observed for the \DCAXY~and \DCAZ~ratios further indicates that the correlations between the transverse and longitudinal DCA components are preserved. Overall, these observations demonstrate that SMOTE improves the statistical balance of the dataset while maintaining the DCA-dependent feature of heavy-flavor decay electrons.

\twocolumngrid

\end{document}